\documentclass[%
 reprint,%
 aip,
]{revtex4-2}

\usepackage{mathtools}
\usepackage{dsfont}
\usepackage{amsmath,amssymb}
\usepackage{stmaryrd}

\usepackage{graphicx}
\usepackage{siunitx}
\usepackage{dcolumn}
\usepackage{bm}

\usepackage[utf8]{inputenc}
\usepackage[T1]{fontenc}
\usepackage{mathptmx}
\usepackage{etoolbox}
\usepackage{url}

\usepackage{multirow}

\usepackage{subcaption}
\DeclareCaptionFormat{groupplotformat}{#1 #3}
\usepackage{xcolor}
\usepackage{tikz}
\usepackage{pgf,pgfplots}
\usetikzlibrary{
arrows,%
backgrounds,%
calc,%
decorations.text,%
decorations.pathreplacing,%
external,%
fadings,%
fit,%
intersections,%
matrix,%
shadings,%
shadows,%
shapes,%
shapes.multipart,%
spy,%
patterns,%
plotmarks,%
through}
\pgfplotsset{compat=1.18} 
\usepgfplotslibrary{groupplots,fillbetween}

\pgfplotsset{
    tuft/.style = {
        width=\columnwidth,
        clip=false,
        axis lines*=left,
        mark size=2,
        axis line shift=15pt,
        tick style={thick,black},
        major tick length=0.15cm,
        x tick label style={rotate=-60,anchor=west}
    }
}

\definecolor{plasmacolor}{RGB}{242,170,250}

\makeatletter
\def\@email#1#2{%
 \endgroup
 \patchcmd{\titleblock@produce}
  {\frontmatter@RRAPformat}
  {\frontmatter@RRAPformat{\produce@RRAP{*#1\href{mailto:#2}{#2}}}\frontmatter@RRAPformat}
  {}{}
}%
\makeatother

\newcommand{\diff}{\mathrm{d}}

\begin{document}

\def\mytitle{A low-temperature plasma collisional framework for quadrature-based moment methods}
\title[\mytitle]{\mytitle}
%%% First author
\author{Pierre-Yves C. R. Taunay}
 \affiliation{Aerospace Engineering Department, United States Naval Academy, Annapolis, MD 21402, USA}
 \email{taunay@usna.edu}

\date{\today}

\begin{abstract}
The mathematical framework for elastic and inelastic collisions for quadrature-based moment methods (QBMM)
is developed and validated for low-temperature collisional plasma physics problems represented by the Boltzmann equation. 
QBMM are more computationally expedient than direct Eulerian solvers and are able
to provide a noise-free solution to the Boltzmann equation, as well as capture non-equilibrium velocity distribution functions
(VDF).
Closure for QBMM is provided by considering that the VDF is a sum of weighted kernel functions placed at discrete velocity abscissas.
For all plasma particles (electrons, ions, and neutrals),
expressions for the reaction integrals within the QBMM framework are derived and generalized to an arbitrary moment order
for elastic collisions represented by a BGK operator 
and inelastic collisions (ionization and excitation) represented by a Boltzmann operator.
The accuracy of different QBMM to compute reaction integrals is discussed, and numerical solutions of the moment model are presented for multiple low-temperature plasma physics problems
that feature both elastic and inelastic collisions. 
It is shown that the Extended Quadrature Method of Moments, which uses a sum of weighted continuous Gaussian kernels, is most advantageous to compute reaction integrals. Good agreement with results from the literature is obtained for the evolution of the VDF and integrated moments, thus demonstrating
that the QBMM collisional framework adequately captures the underlying physics of low-temperature plasmas. 
\end{abstract}

\maketitle

\onecolumngrid
\noindent
\fbox{
\parbox{\textwidth}{
\textbf{The views expressed in this article are those of the author and do not reflect the official policy or position of the U.S. Naval Academy, Department of the Navy, Department of Defense, the U.S. Government, or any agency thereof.\\
\noindent
The following article has been submitted to Physics of Plasmas. \\
\noindent
{Copyright (2026) Pierre-Yves C. R. Taunay. This article is distributed under a Creative Commons Attribution-NonCommercial 4.0 International (CC BY-NC) License. \url{https://creativecommons.org/licenses/by-nc/4.0/}}
}
}
}
\vspace{20pt}
\twocolumngrid

%%%%%%%%%%%%%%%%%%%%%%%%%%%%%%%%%%%%%%%%%%%%%%%%%%%%%%%%%%%%%%%%%%%%%%%%%%%%%%%%%%%%%%%%%%%%%%%%%%%%%%%%%%%%%%%%%%
\section{Introduction}

Low-temperature plasmas feature a variety of non-equilibrium  
effects such as kinetic instabilities or kinetic wave generation due to the presence of collisions between
charged and neutral species, as well as electric and magnetic fields.
As a result, the velocity distribution function (VDF) of electrons is often non-Maxwellian and the
electron temperature differs from that of the heavier species.
Deviations from classical transport theory have been observed in 
devices that feature
strong kinetic effects, such as those with crossed electric and magnetic fields (\textit{e.g.}, Hall-effect thrusters)\cite{Janes1966,Meezan2001} and those with current-driven instabilities (\textit{e.g.}, hollow cathodes).\cite{Mikellides2005,Mikellides2007} 
Strong non-equilibrium also exists in the plasma sheath, owing to the collisionless nature of the plasma in that region. 
The accurate calculation of those kinetic effects is critical
to the understanding of the fundamental physics of 
devices that feature such effects. 

Both 
direct Eulerian solvers\cite{Broadwell1964,Xu2010,Hara2012,Juno2018,Raisanen2019}
and
Lagrangian particle approaches (\textit{e.g.}, particle-in-cell),\cite{Bird1994,Birdsall2004} 
may be used to accurately simulate kinetic effects in plasmas and rarefied gases.
However, both are computationally expensive. 
Direct-discretization solvers suffer from the ``curse of dimensionality''
because they solve the Boltzmann equation in both physical \textit{and} velocity space, for a
total of up to seven dimensions (three dimensions for both physical space and velocity space, and one time dimension).
While Lagrangian approaches avoid this issue by directly
computing particle-particle interactions and
tracking macro-particles, they require both
a large number of particles and floating point operations, especially in higher dimensions, and may also suffer from statistical noise.\cite{Langdon1979,Hara2012,Janhunen2018}

Moment methods are an alternative to both direct-Euler and Lagrangian kinetic solvers for
the simulation of non-equilibrium plasmas. 
Because they rely on the velocity-integrated form of the Boltzmann equation, 
moment methods provide a numerical solution free of statistical noise and are less computationally
expensive than direct-Euler approaches.\cite{Taunay2023}
However, moment methods rely on a ``closure'' assumption so that the set of velocity-integrated
equations can be solved.
Examples of closures include Maxwellian closures where the VDF is assumed to be at or near equilibrium,\cite{Sahu2020}
Grad's closure\cite{Grad1949} 
and its variants,\cite{Struchtrup2003,Torrilhon2004,Kuldinow2024-1,Kuldinow2024-2,Alvarez2022,Alvarez2026}
maximum-entropy closure,\cite{Mead1984,Levermore1996,Boccelli2020} quadrature-based
closures,\cite{McGraw1997,Fox2008,Desjardins2008,Chalons2010,Crestetto2012,Yuan2012,Cheng2014,Nguyen2016,Chalons2017,Pigou2018,Fox2018,Fox2021,Taunay2022,Taunay2023,Fox2023} or other hybrid or data-driven closures.\cite{Bohmer2020,Sadr2021}
Near-equilibrium closures are, in general, not well-suited to resolve strong kinetic effects:
an equilibrium assumption must be made (\textit{i.e.}, the VDF is assumed to be Maxwellian or near-Maxwellian), 
and it is often necessary to rely on semi-empirical approaches
such as anomalous collision models\cite{Mikellides2005,Sary2017,Ortega2018} to accurately represent experimentally observed trends. 
Grad's closures are typically further limited due to the shape of the assumed VDF: in some instances the VDF
may become negative.\cite{Alvarez2022}

This work focuses on Quadrature-based Moment Methods (QBMM). 
Unlike Grad's method, QBMM cannot generate a negative VDF, which is advantageous when computing reaction integrals.
Because they are rooted in linear algebra and do not rely on an optimization routine, QBMM are also computationally advantageous
as compared to maximum-entropy closures. 
In QBMM, the VDF is assumed to be a sum 
of weighted, strictly positive kernel functions placed at discrete velocity abscissas: 
the VDF is the weight function for the inner product associated with (unknown) orthogonal polynomials.
In practice, this means that the weights and abscissas of each QBMM are related to the 
eigenvectors and eigenvalues of the Jacobi matrix that represents the three-term recurrence of these orthogonal polynomials.
{By definition, the orthogonal polynomials, $P_n$, follow a three-term recurrence of the type $P_n\left(X\right) = (A_n X + B_n) P_{n-1}\left(X\right)- C_n P_{n-2}\left(X\right)$, where $A_n$, $B_n$, and $C_n$ are constants, $A_n > 0$, and $C_n > 0$.\cite{Szego}}
Because the weights and abscissas are obtained from sparse, tridiagonal matrices, the method is typically computationally efficient. 

In the context of plasma physics, QBMM have been shown to accurately simulate classical kinetic problems such as collisionless Landau-Damping,\cite{Crestetto2012,Taunay2022} highly collisional plasmas,\cite{Cheng2014} weakly collisional and magnetized plasmas,\cite{Taunay2023} and the ion dynamics of a bounded plasma, both in the bulk plasma and the sheath.\cite{Berger2025}
Collisions relevant to low-temperature plasmas have been implemented by both Taunay and Mueller\cite{Taunay2022,Taunay2023} and Berger~\textit{et al.}.\cite{Berger2025}
However, these results are limited to elastic collisions 
that are represented by a 
Bhatnagar-Gross-Krook (BGK) operator
for a single species\cite{Taunay2022,Taunay2023} 
or for multiple species with a set reaction rate;\cite{Berger2025} 
a more general collisional framework for QBMM remains to be developed.
The goal of this work is to develop this mathematical framework in the context of QBMM to compute the effect of both elastic and inelastic collisions 
in collisional, unmagnetized low-temperature plasmas.
This framework will be developed for the Hyperbolic Quadrature Method of Moments (HyQMOM),\cite{Fox2018,Fox2021,vanCappellen2021}
the Extended Quadrature Method of Moments (EQMOM),\cite{Chalons2010,Yuan2012,Nguyen2016,Chalons2017,Pigou2018} and
the Generalized Quadrature Method of Moments (GQMOM).\cite{Fox2023}
In all cases, because the shape of the VDF in QBMM ensures that a closed-form of the reaction integral exists, 
the mathematical framework is generalized to an arbitrary high moment order. In practice, however, double-precision typically limits
the maximum moment order due to rounding errors and numerical precision.

The underlying numerical scheme and quadrature-based moment methods are briefly described in Section II. 
The collisional framework is delineated in Section III,
and the effect of different QBMM on reaction integrals is explored. 
Finally, results for multiple low-temperature plasma physics problems that feature elastic and inelastic collisions
for multiple species (electrons, ions, and neutrals) are presented and discussed in Section IV. 

%%%%%%%%%%%%%%%%%%%%%%%%%%%%%%%%%%%%%%%%%%%%%%%%%%%%%%%%%%%%%%%%%%%%%%%%%%%%%%%%%%%%%%%%%%%%%%%%%%%%%%%%%%%%%%%%%%
\section{Numerical scheme}
	\subsection{System of equations and assumptions}

    In the absence of a magnetic field, the evolution of the velocity distribution function, $f$, for plasma particles of charge $q$ and mass $m$ is governed by
	the Boltzmann-Poisson system of equations:
    \begin{gather}
			\frac{\partial f}{\partial t}
			+ \mathbf{v}\frac{\partial f}{\partial \mathbf{x}}
			+ \frac{q}{m} 
			\left(
			\mathbf{E}\frac{\partial{f}}{\partial \mathbf{v}}
			\right) 
			= \mathcal{C}\left[f\right],
			\label{eqn:collisional-vlasov-eqn}
			\,\textrm{and}\\
			\nabla^2\phi 
			 = 
			-\frac{1}{\epsilon_0}
			\left(\sum_{s} q_s \int_{\mathds{R}^3} f_s\left(\mathbf{x},\mathbf{v},t\right)\,\mathrm{d}^3\mathbf{v}\right),
			\label{eqn:poisson-equation}
		\end{gather}
where
	$\epsilon_0$ is the permittivity of vacuum,
	$\mathbf{E} = -\nabla\phi$ is the electric field, 
	and $\mathcal{C}$ is a collision operator.
    We assume in this work that the motion of charges does not generate a significant magnetic field
    (electrostatic assumption) and that no external magnetic fields are applied. 
    The index $s$ in Equation~\ref{eqn:poisson-equation} corresponds to the charged species $s$ 
    with charge $q_s$ and VDF $f_s$.

We restrict ourselves to a single physical dimension (1D). 
In some instances (\textit{e.g.}, back-scattering or BGK operators), 
a single velocity dimension is sufficient to describe the plasma physics problem of interest.\cite{Berger2025}
However, because particle scattering is 
three-dimensional for the collisional processes that we will consider (elastic, ionization, excitation), all three velocity dimensions must be kept. 
With those assumptions, the Boltzmann equation becomes
\begin{equation}
\frac{\partial f}{\partial t}
+ v_x\frac{\partial f}{\partial x}
+ \frac{q}{m} 
\left(
E_x\frac{\partial{f}}{\partial v_x}
\right) 
= \mathcal{C}\left[f\right],
\label{eqn:collisional-vlasov-eqn-1d3v}
\end{equation}
where $f$ depends on the physical position ($x$), the three velocity dimensions ($v_x$, $v_y$, and $v_z$), and time ($t$).

The velocity moments of the VDF, $M_{ijk}$, are defined as  
	\begin{equation}
			M_{ijk}\left(x,t\right) = \int_{\mathds{R}^3} v_x^i v_y^j v_z^k
			f\left(x,v_x,v_y,v_z,t\right)\,\mathrm{d}^3\mathbf{v},
			\label{eqn:moment-definition}
		\end{equation}
        where the indices $i$, $j$, and $k$ correspond to 
        the moment order in the $x$, $y$, and $z$ directions,
        respectively.
        The total moment order is $N=i+j+k$.
    The species density, $n$, and drift velocity, $\bar{\mathbf{v}}$, are the moments of order $N=0$ and $N=1$, respectively:
    \begin{gather}
        n = M_{000} = M_0 = \int_{\mathds{R}^3} f\left(x,v_x,v_y,v_z,t\right)\,\mathrm{d}^3\mathbf{v}\\
        M_{100} = \int_{\mathds{R}^3} v_x f\left(x,v_x,v_y,v_z,t\right)\,\mathrm{d}^3\mathbf{v}\\
        M_{010} = \int_{\mathds{R}^3} v_y f\left(x,v_x,v_y,v_z,t\right)\,\mathrm{d}^3\mathbf{v}\\
        M_{001} = \int_{\mathds{R}^3} v_z f\left(x,v_x,v_y,v_z,t\right)\,\mathrm{d}^3\mathbf{v},
    \end{gather}
    such that $\bar{\mathbf{v}} = \left(M_{100}/M_0,M_{010}/M_0,M_{001}/M_0\right)$.
	Integrating Equation~\ref{eqn:collisional-vlasov-eqn-1d3v} multiplied by $v_x^i v_y^j v_z^k$ yields
	an unclosed (\textit{i.e.}, infinite) system of equations for the velocity moments,
	\begin{equation}
			    \frac{\partial M_{ijk}}{\partial t}
			    + \frac{\partial M_{i+1,jk}}{\partial x}
			    = i \frac{q}{m} E M_{i-1,jk} 
			    + \int_{\mathds{R}^3}v_x^i v_y^j v_z^k\mathcal{C}\left[f\right]\,\mathrm{d}^3\mathbf{v},
			\label{eqn:integrated-collisional-vlasov-eqn}
		\end{equation}
	where the PDE for the moment of order $i$ involves moments of higher order (\textit{e.g.}, $i+1$) and 
    a collision operator that may be unclosed.
    
    We further restrict our study to deviations from equilibrium in the 1D direction only (parallel to $x$).
    If the parallel and transverse directions can be decoupled due to the nature of the physical problem (\textit{i.e.}, the problem is 1D1V), the approximate
    VDF, $\tilde{f}$, is:
    \begin{equation}
        \tilde{f}\left(x,v_x,v_y,v_z,t\right) \approx 
        \tilde{f}_{x}\left(v_x\right)
        \delta\left(v_y\right)
        \delta\left(v_z\right),
        \label{eqn:simplifying-1D1V}
    \end{equation}   
    where $\delta$ here denotes the Dirac delta function.
    In general, however, we will assume equilibrium with zero drift in the transverse directions ($y$ and $z$).
    For those cases, $\tilde{f}$ is:
    \begin{equation}
    \begin{aligned}
        \tilde{f}\left(x,v_x,v_y,v_z,t\right) &\approx 
        \tilde{f}_{x}\left(v_x\right) \times \dots \\
        & \dots \left(\dfrac{m}{2 \pi k_B T_\perp}\right) 
        \exp\left(-\dfrac{m}{2 k_B T_\perp}\left(v_y^2 + v_z^2\right)\right)
        .
        \label{eqn:simplifying-transverse-equilibrium}
    \end{aligned}
    \end{equation}
    In those instances, unlike Berger~\textit{et al.}\cite{Berger2025} and Taunay and Mueller,\cite{Taunay2022} we do not fully decouple the parallel and transverse directions
    (\textit{i.e.}, we do not change the 1D3V problem into 1D1V): 
    although the drift velocity in the transverse direction is assumed to be zero (\textit{i.e.}, $M_{010} = M_{001} = 0$), the temperature may be 
    anisotropic (\textit{i.e.}, $T_\perp \neq T_\parallel$), for example in the presence of strong magnetic fields.
    Because we assumed that there are no magnetic fields, we will consider here that the transverse
    temperature is equal to that of the parallel direction: $T_\perp = T_\parallel = T$.
    Because the transverse drift velocity in the simplified VDF is exactly zero and because
    the VDF is assumed to be Maxwellian in the transverse direction,
    all odd-order moments in the transverse direction will vanish, and the 
    second-order moments in the $y$ and $z$ directions are equal: $\forall l \in \mathds{N},\, M_{i,2l+1,k} = M_{ij,2l+1} = 0$, and $M_{020} = M_{002}$. 

    The velocity moments may be further simplified using the approximate VDF, $\tilde{f}$:
    \begin{equation}
       M_{ijk} =  
        G_j G_k
        \int_{-\infty}^{+\infty}
        v_x^i
        \tilde{f}_{x}\left(v_x\right)
        \diff v_x
        = 
        M_i G_j G_k,
    \end{equation}
    where $G_j$ is the moment of order $j$ of a normalized, centered Gaussian distribution with temperature $T_\perp=T$, 
    and $M_i$ is the $i$-th order velocity moment of the $x$-direction VDF.
    By definition,
    \begin{equation}
        G_k = \begin{cases}
            0,\,k\,\textrm{odd}\\
            \left(k-1\right)!! \left(\dfrac{k_B T}{m}\right)^{k/2},\,k\,\textrm{even}.
        \end{cases}
    \end{equation}
    The operator ``$!!$'' is the double factorial.
        
    It is useful to define additional higher-order moments of the VDF that are representative of physical quantities. 
    The temperature, $T$, and heat flux $\mathbf{q} = (q_x,0,0)$, are defined as the moments of the VDF of order two and three 
    with respect to the peculiar velocity,
    $\mathbf{c} = \mathbf{v}-\bar{\mathbf{v}}$: 
    \begin{gather}
        T = \dfrac{1}{3 M_0} \int_{\mathds{R}^3} \left(\mathbf{v} - \bar{\mathbf{v}}\right)^2 f \diff^3\mathbf{v}\\ 
        q_x = \dfrac{1}{2} \int_{\mathds{R}^3} \left(\mathbf{v} - \bar{\mathbf{v}}\right)^2 \left(v_x - \bar{v_x}\right) f \diff^3\mathbf{v},
    \end{gather}
    where $T$ is here in units of specific energy (J/kg) such that the temperature in Kelvin is $T_K =  mT/k_B$ ($m$ here denotes the particle mass),
    and $q_x$ is in units of specific power per unit area (W/kg/m\textsuperscript{2}).
    Both may be recast in terms of the non-central velocity moments:
    \begin{equation}
    \begin{aligned}
        T &=\dfrac{1}{3}\left(T_\parallel + 2 T_\perp\right)\\ 
        &= \dfrac{1}{3 M^2_0}\left( \left(M_{200} + M_{020} + M_{002}\right) M_0 - M^2_{100}\right),\,\textrm{and}
    \end{aligned}
    \end{equation}
    \begin{equation}
    \begin{aligned}
         q_x &= \dfrac{1}{2}\left(M_{300} + M_{120} + M_{102} 
        + 2 \dfrac{M_{100}^3}{M_0^2} \right.\\ 
        &\left. - \dfrac{M_{100}}{M_0}\left(M_{020} + M_{002}\right)
        - 3 \dfrac{M_{100}}{M_0}M_{200}\right)
    \end{aligned}
    \end{equation}
    
	\subsection{Computational implementation}
	In this work, the PDE operator is separated into the transport, collision, and Lorentz force steps with Strang splitting.
	The corresponding differential equations to solve are (for the electrostatic case):
	\begin{align}
		\frac{\partial M_{ijk}}{\partial t} &= -\frac{\partial M_{i+1,jk}}{\partial x},
		\label{eqn:transport-strang}
		\\
		\frac{\textrm{d} M_{ijk}}{\textrm{d}t} 
		&= 
        \int_{\mathds{R}^3}v_x^i v_y^j v_z^k\mathcal{C}\left[f\right]\,\mathrm{d}^3\mathbf{v},\label{eqn:collisional-step-strang}\\
		\frac{\textrm{d} M_{ijk}}{\textrm{d}t} &= i\frac{q}{m} E M_{i-1,jk},
		\label{eqn:Efield-step-strang}
	\end{align}
	respectively.
	This second-order splitting consists in solving Equations~\ref{eqn:collisional-step-strang} (effect of collisions)
	and~\ref{eqn:Efield-step-strang} (effect of electrostatic field) successively over a half time step, then
	Equation~\ref{eqn:transport-strang} (moment transport) over a full time step, then  
	Equations~\ref{eqn:Efield-step-strang} and~\ref{eqn:collisional-step-strang} successively over 
	another half time step.
    The collision operators that appear in Equation~\ref{eqn:collisional-step-strang} are each treated separately:
    Equation~\ref{eqn:collisional-step-strang} is solved for each elastic and inelastic collision operator individually.
    The exact approach chosen to approximate the collision terms is delineated in Section III.
    As in Taunay and Mueller,~\cite{Taunay2023}
    the transport step (Equation~\ref{eqn:transport-strang}) is discretized using a finite volume scheme that is
    integrated in time with a forward Euler scheme, and the two individual 
    ODEs for the collision and electrostatic steps are solved with a simple forward Euler scheme. 
    We use a kinetic flux to compute the numerical flux that appears in the finite volume formulation of the transport step.
    The electric potential is updated with Poisson's equation (Equation~\ref{eqn:poisson-equation}) and the number density of the charged species. 
    The Poisson equation is solved numerically with the GMRES solver as implemented in the HYPRE libary.\cite{HYPRE}
    
    In the parallel direction, moments up to $M_{N}$ may be included with a QBMM closure
    for the moment of order $N+1$.
    The transport step for the parallel moments is:
    \begin{equation}
		\frac{\partial M_{N00}}{\partial t} = -\frac{\partial M_{N+1,00}}{\partial x}
        \Leftrightarrow
		\frac{\partial M_{N}}{\partial t} = -\frac{\partial M_{N+1}}{\partial x}
    \end{equation}
    since $M_{N00} = M_N G^2_0$ and $G_0 = 1$.
    For convenience, we will drop the transverse indices $j$ and $k$ when $j=k=0$ if the subscripts and equations are unambiguous.
    Because we assumed that the VDF is Maxwellian in the transverse direction, 
    all transverse moments are either exactly zero (\textit{e.g.}, $M_{030}=0$) or are
    functions of the second-order transverse moment, $M_{020}$. 

	\subsection{Quadrature methods}

    The closure for the moments in quadrature methods is provided by assuming that the VDF is a sum of weighted kernel functions, $K$,
    that are placed at velocity abscissas, $v_\alpha$, in velocity space. 
    The values of the abscissas and weights, $\omega_\alpha$, depend on the quadrature method chosen and the transported moments.

    QBMM for a multi-dimensional velocity space requires successive approximations of the VDF in each velocity dimension.
    In this case, the weights and abscissas that correspond to the higher velocity dimensions (\textit{e.g.}, $y$-direction) depend on that of  
    the lower velocity dimensions (\textit{e.g.}, $x$-direction). 
    However, because of our simplifying assumption of equilibrium in the transverse directions 
    (Equation~\ref{eqn:simplifying-transverse-equilibrium}), QBMM is only needed 
    to approximate the VDF in the $x$-direction.
    This corresponds to:
	\begin{equation}
		\tilde{f}_x\left(x,v_x,t\right) = 
		\sum_{\alpha=0}^{N-1} 
		\omega_{\alpha}\left(x,t\right) 
		K\left(v_x;v_{\alpha}\left(x,t\right),\mathbf{p}\left(x,t\right)\right), 
	\end{equation}
	where $\omega_{\alpha}$ and $v_{\alpha}$ are unknown weights and abscissas.
    Additional parameters to be determined are denoted by $\mathbf{p}$.
	The algorithm to find all unknowns depends on the chosen kernel function
    and requires at least $2 N$ moments in the $x$-direction to solve for the weights and abscissas.
    Any additional parameters and constraints (\textit{e.g.}, hyperbolicity) that must be
    enforced result in additional moments to be transported.  
     The closure for the second-order transverse moment is provided by simply
    enforcing the condition that $T_\perp = T$. Using the definition of the temperature
    and that constraint, we have
    \begin{equation}
        M_{020} = M_{002} = M_{200} - M^2_{100} M^2_0.
    \end{equation}

	We consider in this work the Hyperbolic Quadrature Method of Moments (HyQMOM),\cite{Fox2018,Fox2021,vanCappellen2021} Generalized Quadrature Method of Moments (GQMOM),\cite{Fox2023} and the Extended Quadrature Method of Moments (EQMOM).
    \cite{Chalons2010,Yuan2012,Nguyen2016,Chalons2017,Pigou2018} 
	Implementation details for HyQMOM and EQMOM may be found in Ref.~\onlinecite{Taunay2023}, and in Ref.~\onlinecite{Fox2023} for GQMOM.
    Because each QBMM has distinct advantages (\textit{e.g.}, continuous kernels for EQMOM or computational expediency for HyQMOM), different QBMM may be used at any given stage of the numerical solution. 
    We will use HyQMOM to transport moments as it has been shown
    to do so 
    accurately in the context of plasma physics.\cite{Taunay2022,Taunay2023,Berger2025}
    GQMOM and EQMOM, on the other hand, 
    will only be used to compute collision integrals. GQMOM provides a much larger number
    of quadrature nodes to perform collision integrals as compared to HyQMOM, and EQMOM leverages continuous exponential kernels 
    for which a closed form of the collision integrals is known. 
    The accuracy of each method to compute the collision integrals is compared in Section III.

	\subsubsection{Hyperbolic Quadrature Method of Moments (HyQMOM)}
	HyQMOM is an extension of the original quadrature method of moments (QMOM)~\cite{McGraw1997,Fox2008,Desjardins2008} that enforces
	hyperbolicity for the transport equation. Much like QMOM, the kernel function is a Dirac delta function:
	\begin{equation}
		K\left(v;v_\alpha\right) = \delta\left(v-v_\alpha\right).
	\end{equation}
	A total of $2N$+1 moments is required to solve for the weights and abscissas and enforce hyperbolicity.
    The weights and abscissas may be obtained from the 
    the diagonalization of the Jacobi matrix
    of the orthogonal polynomials associated with the quadrature rule.
    The tridiagonal Jacobi matrix is given by
    \begin{equation}
        J_N = \begin{bmatrix}
        \alpha_0 &  \sqrt{\beta_1}  &                       & &  \\
        \sqrt{\beta_1} &   \alpha_1 & \sqrt{\beta_2}                & & \\
               &            & \ddots & &   \\
                &            &                        \sqrt{\beta_{N-1}} & \alpha_{N-1} \\
        \end{bmatrix}
        \label{eqn:hyqmom-jacobi-matrix}
    \end{equation}
    where $\alpha_k$ and $\beta_k$ are computed using the Chebyshev algorithm, with the following recurrence formula:
    \begin{gather}
    \alpha_k = \dfrac{\xi_{k,k+1}}{\xi_{k,k}} - \dfrac{\xi_{k-1,k}}{\xi_{k-1,k-1}}\,\textrm{and}\\ 
    \beta_k = \dfrac{\xi_{k,k}}{\xi_{k-1,k-1}},    
    \end{gather}
    where
    \begin{equation}
        \xi_{k,l} = \xi_{k-1,l+1} - \alpha_{k-1}\xi_{k-1,l}-\beta_{k-1}\xi_{k-2,l}.
    \end{equation}
    The indices $k$ and $l$ are defined in the intervals $k\in \llbracket 0\cdots N-1 \rrbracket$ and $l \in \llbracket k\cdots 2N-1\rrbracket$, respectively.
    The recurrence is initialized with the following conditions:
    \begin{gather}
        \alpha_0 = M_1/M_0, \\
        \beta_0 = M_0, \\
        \xi_{-1,l} = 0,\,\textrm{for}\, l \in \llbracket 1\cdots 2N-2\rrbracket,\,\textrm{and}\\
        \xi_{0,l} = M_l\,\textrm{for}\, l \in \llbracket 0\cdots 2N-1\rrbracket.
    \end{gather}
    Hyperbolicity for the moment transport is enforced in HyQMOM with the following constraints:\cite{Fox2021} 
    \begin{gather}
        \alpha_{N-1} = \dfrac{1}{N-1}\sum_{i=0}^{N-2} \alpha_i,\,\textrm{and}\\
        \beta_{N-1} = \dfrac{2N-1}{N-1}\beta^*_{N-1},
    \end{gather}
    where $\beta^*_{N-1}$ is that obtained from the Chebyshev algorithm.

	\subsubsection{Extended Quadrature Method of Moments (EQMOM)}
Representing the VDF as a discrete set of delta functions
is a clear disadvantage of HyQMOM.
EQMOM~\cite{Chalons2010,Yuan2012,Nguyen2016,Chalons2017,Pigou2018} can be used 
if a continuous representation in velocity space of the VDF is desired. 
{In the context of plasma physics, this is advantageous for computing reaction
	rates for reactions that involve the tail of the VDF. For example, ionization
	events in low-temperature plasmas involve electrons with a much higher energy 
	($\mathcal{O}\left(10^1\right)$~eV) than the bulk electron temperature
	($\mathcal{O}\left(10^0\right)$~eV).}
	We illustrate this particular example in a 1D1V context in Section III.C.2.

A variety of kernels may be considered depending on the support required for the VDF.\cite{Cheng2014,Chalons2010,Chalons2017,Pigou2018,Yuan2012,Nguyen2016}
We consider Gaussian kernels because \textit{(i)} the particle velocity is defined on the 
entire real line and \textit{(ii)} a single Gaussian kernel would correspond to the equilibrium 
condition (\textit{i.e.}, a Maxwellian distribution). 
{This provides a physically consistent reconstruction in the equilibrium limit.}
The Gaussian kernel is given by:
\begin{equation}
	K_G\left(v;v_\alpha,\sigma\right) = \frac{1}{\sqrt{2\pi\sigma^2}}
	\exp\left(-\frac{\left(v-v_\alpha\right)^2}{2\sigma^2}\right),
	\label{eqn:gaussian-kernel}
\end{equation}
where $\sigma$ is the standard deviation that is identical for all $\alpha$.
The corresponding VDF is:
	\begin{equation}
	\tilde{f}\left(v\right) = 
	\sum_{\alpha}
\frac{\omega_{\alpha}}{\sqrt{2\pi\sigma^2}}
\exp\left(-\frac{\left(v-v_\alpha\right)^2}{2\sigma^2}\right)
\end{equation}
The total number of moments
required is $2N+1$ because of the additional unknown (standard deviation, $\sigma$). 
The algorithm to find the variance from an input set of moments is detailed in Ref.~\onlinecite{Taunay2023}.

    \subsubsection{Generalized Quadrature Method of Moments (GQMOM)}
    Much like other Gaussian quadratures, the numerical accuracy of source integrals (\textit{e.g.}, collision integrals) that involve the VDF 
    benefits from additional quadrature points (\textit{i.e.}, nodes).
    However, the total number of nodes (\textit{i.e.}, number of weights and abscissas) 
    is limited in all quadrature-based moment methods 
    by the total number of transported moments,
    and an increasing number of transported moments is not always desirable: transporting
    a moment of order $k$ results in the use of the power function of power up to $k+1$,
    which may run into double- precision limitations.

    To overcome this limitation, Fox~\textit{et al.} introduced 
    the Generalized Quadrature Method of Moments (GQMOM)
    which is
    an extension of the quadrature representation of a QBMM that combines 
    the use of the recursion coefficients of the orthogonal polynomials
    with a free parameter,
    $\nu$, that controls the tail of the VDF.\cite{Fox2023}
    Because an assumption about the shape
    of the tail is made and because the method exactly reconstructs the transported moments, 
    no additional information about the VDF itself is obtained: 
    quadrature points are only added to compute the source integrals. 
    
    GQMOM extends the Jacobi matrix (Equation~\ref{eqn:hyqmom-jacobi-matrix}) with additional quadrature points.
    For $i\leq N-1$, the coefficients $\alpha_k$ and $\beta_k$ are computed with the Chebyshev algorithm 
    delineated in the previous section.
    For $i\geq N$, the coefficients are given by:\cite{Fox2023}
    \begin{gather}
    \alpha_i = \alpha_{N-1},\,\textrm{and}\\
    \beta_i = \left(\dfrac{i}{N-1}\right)^\nu \beta_{N-1}.
    \end{gather}
    Fox~\textit{et al.} indicate that values of $\nu$ of 0, 1, and 2 correspond to ``tails that decay very quickly in a bounded interval,'' 
    ``tails of a Gaussian distribution,'' and ``exponential tails,'' respectively.\cite{Fox2023} 
    While Fox~\textit{et al.} demonstrate that the values of 0, 1, and 2 are intuitively linked to known distributions, the choice of $\nu$ is 
    not limited to those three values only. 
    Unfortunately, no criterion is given
    to choose a value of $\nu$ that allows for the computation of collision integrals. 
    Typically, an increased value of $\nu$ contributes to larger off-diagonal components for the Jacobi matrix.
    This, in turn, corresponds to an increase in the distance between eigenvalues (\textit{i.e.}, abscissas)
    and a decrease in the quadrature weights.
    Because the quadrature weights are normalized such that their sum is always equal to one, the abscissas that
    are placed at increasingly large velocities will have a vanishingly small weight.
    Consider, for example, a two-node Gauss-Hermite quadrature 
    that is extended with GQMOM with one additional node.
    The recurrence coefficients for the Gauss-Hermite quadrature
    are $\alpha_n = 0$ and $\beta_n=n/2$, respectively.\cite{NIST:DLMF}
    The corresponding 2$\times$2 Jacobi matrix ($N=2$) associated with the
    monic, orthonormal version of the Hermite ``physicist'' polynomial is:
    \begin{equation}
        J_2 = \dfrac{1}{\sqrt{2}}
        \begin{bmatrix}
            0 & 1 \\
            1 & 0 
        \end{bmatrix}.
    \end{equation}
    With the GQMOM algorithm, the extended Jacobi matrix is:
    \begin{equation}
        J_3\left(\nu\right)= 
        \dfrac{1}{\sqrt{2}}
        \begin{bmatrix}
            0 & 1 & 0 \\
            1 & 0 & \sqrt{2^\nu} \\
            0 & \sqrt{2^\nu} & 0
        \end{bmatrix},
    \end{equation}
    such that the new off-diagonal component in the Jacobi matrix, $\beta_2$, is:
    \begin{equation}
        \beta_2= \left(2/1\right)^\nu \cdot \beta_1 = 2^{\nu-1}.
    \end{equation}
    The Gauss-Hermite quadrature abscissas are retrieved for the choice of $\nu=1$ in this particular example.
    The corresponding set of abscissas (\textit{i.e.}, eigenvalues) 
    and weights 
    for this example are
    $\left\{0,\pm\sqrt{\left(2^\nu +1\right)/2}\right\}$
       and 
    $1/\left(2+4^\nu\right)\left\{4^\nu,1,1\right\}$, respectively.
    \footnote{To retrieve exactly the Gauss-Hermite quadrature, the weights must be further multiplied by the integral of the weight function: $\int_{-\infty}^{+\infty} \exp\left(v^2\right)\diff{v}$} 
    The largest abscissa is placed at increasingly large 
    velocities ($v_\alpha \sim 2^{\left(\nu-1\right)/2}$ for large $\nu$) with
    a vanishingly small weight ($\omega_\alpha \sim 4^{-\nu})$,
    which may be desirable for threshold processes that have a high-energy cutoff (\textit{e.g.}, ionization).
    
    In general, the numerical accuracy of the source integrals is dependent on the choice of $\nu$.
    However, the optimal value of $\nu$ (\textit{i.e.}, a value of $\nu$ such that the numerical accuracy of the source
    integrals is maximized) depends on the shape of integrand in the source integrals, and,
    therefore, on the shape of both the collision operator considered as well as the underlying (unknown) VDF.
    We illustrate the sensitivity of the source integral on the value of $\nu$ in Section III.C.3 with a pair of 1D1V examples 
    wherein we compute the ionization rate of an argon gas using multiple quadrature methods. 
    To the author's knowledge, no clear criteria to choose $\nu$ have been proposed, and $\nu$ may take any values beyond the three qualitative choices discussed by Fox~\textit{et al.}\cite{Fox2023} The optimal choice of $\nu$ remains
    outside of the scope of this manuscript.

%%%%%%%%%%%%%%%%%%%%%%%%%%%%%%%%%%%%%%%%%%%%%%%%%%%%%%%%%%%%%%%%%%%%%%%%%%%%%%%%%%%%%%%%%%%%%%%%%%%%%%%%%%%%%%%%%%
\section{Collision operators}

    We derive in this section the moments of the Boltzmann operator 
    and of the Bhatnagar-Gross-Krook (BGK)\cite{Bhatnagar1954} collision operator 
    and discuss their implementation within the QBMM framework.
    The BGK and Boltzmann operators will be used to model elastic and inelastic collisions, respectively.
    Although the use of the BGK operator is strictly valid near thermodynamic equilibrium,\cite{Haack2017,Boccelli2020,Alvarez2026}
    we demonstrate that it is sufficient for the applications considered herein
    and explore the limitations of this approach.
    The more general Boltzmann operator considered for inelastic collisions can be extended in a similar
    fashion to elastic collisions.

\subsection{Elastic collisions: Bhatnagar-Gross-Krook operator}

    The BGK operator must conserve the first three total moments
	(mass, momentum, and energy) of the two species involved in the collision.
	For species $i$, the collision operator for an elastic collision between species $i$ and $j$ is given by:
	\begin{equation}
		\left(\dfrac{\partial f_i}{\partial t}\right)_\textrm{elastic} \approx \mathcal{C}_{ij,\textrm{BGK}} = \nu_{ij} \left(\mathcal{M}_{ij} - f_i\right),
	\end{equation}
	where $\nu_{ij}$ is the collision frequency. 
	The distribution $\mathcal{M}_{ij}$ is a Maxwellian distribution function with a cross-species drift velocity, $\mathbf{u}_{ij}$,
	and temperature, $T_{ij}$:
	\begin{equation}
		\mathcal{M}_{ij}\left(\mathbf{v}\right) = n_i \left( \dfrac{m_i}{2 \pi T_{ij}}\right)^{d_V/2} 
		\exp\left(-\dfrac{m_i \left(\mathbf{v}-\mathbf{u}_{ij}\right)^2}{2 T_{ij}}\right),
	\end{equation}
	where $d_V$ is the number of dimensions considered in velocity space (\textit{i.e}, $d_V=1$ for a 1D1V problem)
	and the temperature is in units of energy.
    The cross-species quantities depend on the mass, drift velocity, and temperature of both species $i$ and $j$.
    The method by which the cross-species quantities, $\mathbf{u}_{ij}$ and $T_{ij}$, are computed is non-unique and may affect
    numerical results.
	We follow the framework from Haack~\textit{et al.}\cite{Haack2017} to ensure that the elastic collision
	operator conserves mass, momentum, and energy, and satisfies the Boltzmann $\mathcal{H}$-theorem.
	Conservation laws for the first three moments may be written as:
	\begin{align}
		\int  \mathcal{C}_{ij} \diff\mathbf{v} &= 0 \label{eqn:bgk-mass-conservation}\\
		\int  m_i \mathbf{v} \mathcal{C}_{ij} \diff\mathbf{v} 
			+ \int m_j \mathbf{v} \mathcal{C}_{ji} \diff\mathbf{v} &= 0,\,\textrm{and} \label{eqn:bgk-momentum-conservation}\\
		\int  \dfrac{1}{2} m_i \mathbf{v}^2 \mathcal{C}_{ij} \diff\mathbf{v} 
		+\int  \dfrac{1}{2} m_j \mathbf{v}^2 \mathcal{C}_{ji} \diff\mathbf{v} &= 0,\label{eqn:bgk-energy-conservation}
	\end{align}
	respectively.
	Equation \ref{eqn:bgk-mass-conservation} is automatically enforced due to the definition of $\mathcal{M}_{ij}$.
	Conservation laws for momentum and energy (Equations~\ref{eqn:bgk-momentum-conservation} and \ref{eqn:bgk-energy-conservation})
	yield a constraint for the cross-species drift velocity and temperature, respectively:
	\begin{equation}
		\mathbf{u}_{ij} = \dfrac{m_i n_i \nu_{ij} \mathbf{u}_i + m_j n_j \nu_{ji} \mathbf{u}_j}{m_i n_i \nu_{ij}+m_j n_j \nu_{ji}},\,\textrm{and}
		\label{eqn:bgk-cross-drift}
	\end{equation}
    \begin{equation}
    \begin{split}
		{T}_{ij} &= 
		\dfrac{n_i \nu_{ij} T_i + n_j \nu_{ji} T_j}{n_i \nu_{ij} + n_j \nu_{ji}} + \dots\\
		& \dots\dfrac{m_i n_i \nu_{ij} \left(\mathbf{u}_i^2 - \mathbf{u}_{ij}^2\right) 
			+ m_j n_j \nu_{ji}\left(\mathbf{u}_j^2-\mathbf{u}_{ij}^2\right)}
			{d_V\left(n_i \nu_{ij}+n_j \nu_{ji}\right)}
		\label{eqn:bgk-cross-temperature},
    \end{split}
	\end{equation}
	where $\mathbf{u}_i$ is the drift velocity (\textit{i.e.}, the first-order moment) of species $i$.
	
	For a BGK operator, the collision frequency may be computed --- unless otherwise specified ---using the definition of the reaction rate, $R_{ij}$:
\begin{equation}
	\nu_{ij} = \dfrac{R_{ij}}{n_i}.
	\label{eqn:collision-rate-from-reaction-rate}
\end{equation}
The reaction rate is given by:
\begin{equation}
	R_{ij} = \iint \left|\mathbf{v}_i - \mathbf{v}_j\right| 
    Q\left(\left|\mathbf{v}_i - \mathbf{v}_j\right|\right) f_i f_j
	\diff^3 \mathbf{v}_i \diff^3\mathbf{v}_j,
	\label{eqn:reaction-rate-generic}
\end{equation}
where $Q$ is the (total) cross section for the type of collision considered
and $\mathbf{v}_i$ is the velocity of species $i$.
We also enforce the condition $m_i n_i \nu_{ij} = m_j n_j \nu_{ji}$.
	The cross-species velocity and temperature are then given by \cite{Haack2017}:
	\begin{align}
		\mathbf{u}_{ij} &= \dfrac{\mathbf{u}_i + \mathbf{u}_j}{2},\,\textrm{and}\\
		{T}_{ij} &= 
		\dfrac{m_j T_i +m_i T_j}{m_i+m_j}
		+ \dfrac{1}{2 d_V}\dfrac{m_i m_j}{m_i+m_j} 
		\left(\mathbf{u}_i-\mathbf{u}_{j}\right)^2,
	\end{align}	
	respectively. The term $\left(\mathbf{u}_i-\mathbf{u}_{j}\right)^2$ corresponds to the square of the magnitude of the vector
    $\mathbf{u}_i - \mathbf{u}_j$.

	\subsubsection{Electron-ion collisions}
	\label{sec:electron-ion-collisions}
    For the BGK operator, the system of equations for the evolution of the electron and ion VDF, $f_e$ and $f_i$, due to electron-ion collisions is:\cite{NRLFormulary}
	\begin{align}
		\dfrac{\partial f_e}{\partial t} &= \nu_{ee} \left(\mathcal{M}_{ee} - f_e\right) + \nu_{ei} \left(\mathcal{M}_{ei} - f_e\right)
		\label{eqn:coulomb-collisions-electron}
		\\
		\dfrac{\partial f_i}{\partial t} &= \nu_{ie} \left(\mathcal{M}_{ie} - f_i\right) + \nu_{ii} \left(\mathcal{M}_{ii} - f_i\right)
		\label{eqn:coulomb-collisions-ion},
	\end{align}
	where $\mathcal{M}_{ee}$ and $\mathcal{M}_{ii}$ are Maxwellian VDFs which have the same first three moments as the electron and ion VDF, respectively. 
	The Maxwellian VDF for species $s$ is given by:
	\begin{equation}
		\mathcal{M}_{ss}\left(\mathbf{v}\right) = n_s \left( \dfrac{m_s}{2 \pi T_{s}}\right)^{d_V/2} 
\exp\left(-\dfrac{m_s \left(\mathbf{v}-\mathbf{u}_{s}\right)^2}{2 T_{s}}\right).
	\end{equation}
	The solution of this system of equations for species $s$ colliding with species $r$ over the interval $\left[0,t\right]$ is:
	\begin{equation}
    \begin{split}
		f_s\left(t\right) &= f_{s,0} \exp\left(-t \nu_T\right) + \dots \\
        & \dots \dfrac{1-\exp\left(-t \nu_T\right)}{\nu_T} \left(\nu_{sr} \mathcal{M}_{sr} - \nu_{ss}\mathcal{M}_{ss}\right),
        \label{eqn:electron-ion-vdf-solution}
    \end{split}
	\end{equation}	
	where $f_{s,0}$ is the initial VDF and $\nu_{T} = \nu_{sr} + \nu_{ss}$.
    Here, $\nu_{ss}$ denotes the self-collision frequency, and $\nu_{sr}$ the cross-species collision frequency.
	Equation~\ref{eqn:electron-ion-vdf-solution} may be integrated over the velocity space to obtain an analytical solution
	for the moments of order $l=i+j+k$ (\textit{i.e.}, $M_{ijk}$) of species $s$, $M_{l,s}$:
	\begin{equation}
    \begin{split}
	M_{l,s}\left(t\right) &= M_{l,s}\left(t=0\right) \exp\left(-t \nu_T\right) + \dots \\
    &
    \dots
	\dfrac{1-\exp\left(-t \nu_T\right)}{\nu_T} 
	\left(\nu_{sr} \bar{M}_{l,sr} - \nu_{ss}\bar{M}_{l,ss}\right),
    \end{split}
	\end{equation}		
	where $\bar{M}_{k}$ indicates the moment of order $l$ of a Maxwellian.
	We use the collision frequency for Coulomb interactions for the momentum relaxation rate, as given by Ref.~\onlinecite{Haack2017} (in CGS units):
	\begin{equation}
		\nu_{ij} = \dfrac{1}{m_i n_i} 
		\dfrac{8 n_i n_j \sqrt{2\pi m_i m_j} \left(Z_i Z_j e^2\right)^2 \left(m_i+m_j\right)}
		{3 \left(m_i T_j + m_j T_i\right)^{3/2}} \Lambda_{ij}
	\end{equation}
	where $\Lambda_{ij}$ is the Coulomb logarithm, $Z_i$ is the ionization state of species $i$, and $e$ is the elementary charge.
	The Coulomb logarithm for each species interaction is given by (in CGS units):\cite{NRLFormulary}
	\begin{align}
		\Lambda_{ei} &= 23 - \dfrac{1}{2} \ln\left( \dfrac{n_e}{T_{e,eV}^3}\right),\\
		\Lambda_{ii} &=  23 - \dfrac{1}{2} \ln\left( 2 \dfrac{n_i}{T_{i,eV}^3}\right),\textrm{and}\\
        \begin{split}
		\Lambda_{ee} &= 23.5 
		- \dfrac{1}{2} \ln\left(n_e\right) 
		+ \dfrac{5}{4} \ln\left(T_{e,eV}\right) - \dots\\
		&\dots \left(10^{-5} + \dfrac{\left(\ln\left(T_{e,eV}\right)-2\right)^2}{16}\right)^{1/2}.
        \end{split}
	\end{align}

	\subsubsection{Electron-neutral and ion-neutral collisions}
	The system of equations for the evolution of the electron (or ions) and neutral VDF, $f_e$ (resp. $f_i$) and $f_n$, due to elastic collisions is:
	\begin{align}
		\dfrac{\partial f_e}{\partial t} &= \nu_{en} \left(\mathcal{M}_{en} - f_e\right)\\
		\dfrac{\partial f_n}{\partial t} &= \nu_{ne} \left(\mathcal{M}_{ne} - f_n\right).
	\end{align}
	Each ODE may be independently solved. The solution for species $s$ over the interval $\left[0,t\right]$ is:
	\begin{equation}
		f_s\left(t\right) = f_{s,0} \exp\left(-t \nu_{sr}\right)+ \left(1-\exp\left(-t  \nu_{sr}\right)\right)
		\mathcal{M}_{sr}.
		\label{eqn:electron-neutral-vdf-solution}
	\end{equation}		
	The procedure outlined in Section~\ref{sec:electron-ion-collisions} may be repeated to obtain an analytical solution
	for the moments of order $k$ of species $s$, $M_{k,s}$:
\begin{align}
	M_{k,s}\left(t\right) &= M_{k,s}\left(t=0\right) \exp\left(-t \nu_{sr}\right)+\dots\\ 
	&\dots 
    \left(1-\exp\left(-t \nu_{sr}\right)\right)
	\bar{M}_{k,sr}.
\end{align}		
    
    The neutrals are not updated in the test cases considered in this manuscript: $\forall t,\, T_{n} = T_n\left(t=0\right)$, and $\mathbf{u}_n = \mathbf{0}$. 
    For elastic electron-neutral collisions, we also have $m_e \ll m_n$ such that the cross-species 
    terms in the relaxed Maxwellian distribution $\mathcal{M}_{en}$ are $\mathbf{u}_{en} = \dfrac{1}{2} \mathbf{u}_e$ and 
    $T_{ij} \approx T_e + \dfrac{1}{6} m_e \mathbf{u}^2_e$. 
    This, in turn, means that there exists an external mechanism to keep the neutral energy 
    constant at the expense of the electron energy. This inconsistency can be remedied by tracking the
    moments of the neutrals as well but is beyond the scope of this manuscript.

\subsection{Inelastic collisions: Boltzmann operator}

\subsubsection{Moments of collision operator}
We follow a framework similar to that of Alvarez-Laguna~\textit{et al.}\cite{Alvarez2022,Alvarez2026} and derive the moments
of the collision operator for the electron and heavy species.
The Boltzmann operator for collisions between particles of type $i$ and $j$, is given by:\cite{Alvarez2022}
\begin{equation}
	\mathcal{C}_{ij} = \int_{\mathds{R}^3}\int_{\mathds{S}^2}
	g
	\left(f'_i f'_j - f_i f_j\right)
	\sigma\left(g,\Omega\right) 
	\diff \Omega \diff^3\mathbf{v}_j,		
\end{equation}
where primed quantities are that after the collision, $\sigma$ is the differential cross section, $\diff\Omega$ is the
solid angle of the post-collision particle in the center-of-mass frame of reference, and 
$g = \left|\mathbf{g}\right| = \left|\mathbf{v}_i - \mathbf{v}_j\right|$.
The $k$-th-order moment of the collision operator for particle $i$ is given by:
\begin{equation} 
	M^i\left[\mathcal{C}_{ij}\right]_{k} = \int \psi_k \mathcal{C}_{ij} \diff^3\mathbf{v}_i,
	\label{eqn:moment-cx-operator-general}
\end{equation}
where $\psi_k = \mathbf{v}_i^k$. 
Our definition of $\psi_k$ differs from that of Alvarez-Laguna~\textit{et al.}: we use the velocity of species $i$
as opposed to the peculiar velocity, as we are interested in the evolution of the moments of the VDF, as opposed to
the central moments.
Because $\psi_k = \mathbf{v}_i^k = \left(v_x,v_y,v_z\right)^k$, the time-derivative term in the corresponding inelastic collision differential 
equation (Equation~\ref{eqn:collisional-step-strang})
is a multinomial expansion:
\begin{equation}
\begin{split}
M^i\left[\mathcal{C}_{ij}\right]_{k} &= \int \psi_k \mathcal{C}_{ij} \diff^3\mathbf{v}_i \\
&= 
    \frac{\textrm{d}}{\textrm{d}t} 
    \left( 
    \sum_{\substack{n_1, n_2, n_3 \geq 0\\n_1+n_2+n_3 = k}}
   \dfrac{k!}{n_1!n_2!n_3!} M_{2n_1,2n_2,2n_3} 
    \right). 
    \label{eqn:inelastic-moments-multinomial}
\end{split}
\end{equation}
Using the approximated VDF, $\tilde{f}$, $M_{2n_1,2n_2,2n_3} = M_{2n_1}G_{2n_2}G_{2n_3}$. 
Equation~\ref{eqn:moment-cx-operator-general} may also be written as:\cite{Alvarez2022}
\begin{equation}
\begin{split}
	&M^i\left[\mathcal{C}_{ij}\right]_{k}
	=\\
	&\int_{\mathds{R}^3}
	\int_{\mathds{R}^3}
	\int_{0}^{\pi}
	\int_{0}^{2\pi}
	g
	\left(\psi'_k - \psi_k\right)
	f_i f_j
	\sigma\left(g,\chi\right) 
	\sin\chi \diff\chi\diff\phi \diff^3\mathbf{v}_i\diff^3\mathbf{v}_j,
    \end{split}
	\label{eqn:moments-of-boltzmann-operator}
\end{equation}
where we used the definition of the solid angle: $\diff\Omega = \sin\chi \diff\chi\diff\phi$.
The angles $\chi$ and $\phi$ are the polar and azimuthal angle of the scattered species post-collision. 

The integrals in Equation~\ref{eqn:moments-of-boltzmann-operator} may be carried out by performing the following change of variables:
\begin{align}
	\mathbf{G} &= \dfrac{m_i \mathbf{v}_i + m_j \mathbf{v}_j}{m_i+m_j}\\
	\mathbf{g} &= \mathbf{v}_i - \mathbf{v}_j.
\end{align}
By conservation of momentum, we have $\mathbf{G} = \mathbf{G}'$.
The pre- and post-collision velocities may be written in terms of the relative and center of mass velocity:\cite{Alvarez2022,Alvarez2026,Zhdanov,Vincenti}
\begin{align}
	\mathbf{v}_i &= \mathbf{G} + \dfrac{\mu}{m_i} \mathbf{g}\\
	\mathbf{v}_j &= \mathbf{G} - \dfrac{\mu}{m_j} \mathbf{g}\\
	\mathbf{v}'_i &= \mathbf{G} + \dfrac{\mu}{m_i} \mathbf{g}'\\
	\mathbf{v}'_j &= \mathbf{G} - \dfrac{\mu}{m_j} \mathbf{g}',
\end{align}
where $\mu = m_i m_j/ (m_i+m_j)$.
By definition, $\diff^3\mathbf{v}_j\diff^3\mathbf{v}_i = \diff^3\mathbf{G}\diff^3\mathbf{g}$.

The pre- and post-collision relative velocities are related through:\cite{Zhdanov}
\begin{equation}
	\mathbf{g}' = g \cos\chi \mathbf{i} + g \sin\chi \cos\phi\,\mathbf{j} + g \sin\chi \sin\phi\,\mathbf{k},
\end{equation}
and
\begin{equation}
	\mathbf{g}' = g \cos\chi \mathbf{i} \dfrac{\left|\mathbf{g'}\right|}{\left|\mathbf{g}\right|} + g \sin\chi \cos\phi\,\mathbf{j} + g \sin\chi \sin\phi\,\mathbf{k},
\end{equation}
for elastic and inelastic collisions, respectively.
The unit vector $\mathbf{i}$ is directed along $\mathbf{g}$ (\textit{i.e.}, $\mathbf{i} = \mathbf{g}/g$) and $\mathbf{j}$ and $\mathbf{k}$ are perpendicular to $\mathbf{i}$.
For inelastic collisions, $\left|\mathbf{g}\right| \neq \left|\mathbf{g}'\right|$.
All the terms parallel to $\mathbf{j}$ and $\mathbf{k}$ involve $\cos\phi$ and $\sin\phi$,
and, therefore, disappear after carrying the integration over $\phi$.
\footnote{Vincenti and Kruger\cite{Vincenti} propose a different parameterization between 
	$\mathbf{g}$ and $\mathbf{g}'$. However, the approaches from either Vincenti and Kruger or Zhdanov 
    yield the same results,
	as all terms involving the azimuthal angle also vanish 
	after integrating over the solid angle.
}

We only consider excitation and impact-ionization inelastic collisions of electrons with neutrals:
we assume that electrons interact with ions only through Coulomb collisions and neglect any electron-ion, ion-ion, and ion-neutral inelastic collisions. 
Conservation of energy yields a constraint on the magnitude of the relative velocities.
For excitation (e + n $\rightarrow$ e + n\textsuperscript{*}), we have:
\begin{equation}
	\dfrac{\mu}{2}{g}^2 =\dfrac{\mu}{2}{g'}^2 + e \phi_{ex},
\end{equation}
where $\phi_{ex}$ is the excitation potential in eV. 
An ionization reaction creates an additional electron, e\textsubscript{2}, with an energy that may differ from that of the
primary electron e\textsubscript{1}: e\textsubscript{1}+n $\rightarrow$ e\textsubscript{1}+ e\textsubscript{2} + i.
Two assumptions may be made:
\begin{itemize}
	\item Equal-sharing: both electrons (e\textsubscript{1} and e\textsubscript{2}) have the same energy
	\item Zero-sharing: the secondary electron has zero energy, while the primary electron has a non-zero energy.
\end{itemize}
For simplicity and ease of comparison to other benchmarks, we assume equal-sharing. The conservation of energy for ionization
is then given by
\begin{equation}
	\dfrac{\mu}{2}{g}^2 =\mu{g'}^2 + e \phi_{iz},	
\end{equation}
where $\phi_{iz}$ is the ionization potential in eV.

The inner integral may now be computed for all $\psi_k$. 
For the plasmas considered, $T_n \ll 1$~eV and $f_n \rightarrow 0$ when $\left|\mathbf{v}_n\right| \approx $ 1--2~eV.
Because the inelastic processes considered have a threshold energy $\phi \sim \mathcal{O}\left(10\right)$~eV,
we then have $\mathbf{g}\approx \mathbf{v}_e$. 
Because $m \ll M$, we also have $\mu\approx m$ and $\mathbf{G} \approx \mathbf{v}_n$, and, therefore, $\mathbf{G}^2 \ll \phi$.
We further neglect all momentum exchange contributions due to ionization and excitation collisions as compared
to elastic collisions and assume isotropic scattering.
With those approximations, Equation~\ref{eqn:moments-of-boltzmann-operator} for the first seven moments ($M_0$ through $M_6$) of excitation reactions becomes:
\begin{equation}
M^{e}\left[\mathcal{C}_{en,ex}\right]
=	 \begin{bmatrix}
		0\\
		0\\
		-2\bar{\phi} R^{(1)}\\
		0 \\
		-4\bar{\phi} R^{(2)} + 4\bar{\phi}^2 R^{(1)}\\
		0\\
		-6\bar{\phi} R^{(3)} + 12\bar{\phi}^2 R^{(2)} - 8\bar{\phi}^3 R^{(1)}
	\end{bmatrix},
\end{equation}
where $\bar{\phi} =  {e\phi}/{m}$ and $R^{(1)}$ is the reaction rate as defined in Equation~\ref{eqn:reaction-rate-generic} with $\left|\mathbf{v}_e-\mathbf{v}_n\right|\approx \left|\mathbf{v}_e\right|$. 
The reaction integral $R^{(k)}$ ($k\geq1$) is defined as:
\begin{equation}
\begin{split}
	R^{(k)} &= n_n 
	\int_{\mathds{R}^3}
	\left|\mathbf{v}_e\right|^{2k-1}
	f_e
	Q^T\left(\left|\mathbf{v}_e\right|\right)
	\diff^3\mathbf{v}_e,
\end{split}
	\label{eqn:reaction-rate-inelastic}
\end{equation}
 where $Q^T$ is the total inelastic collision cross section evaluated at the differential velocity $g$. 
By inspection, the individual coefficients in the $k$-th moment of the collision operator follow OEIS sequence A276985.\cite{A276985}
Therefore, for a moment of arbitrary even order (with $k>0$):
\begin{equation}
	M^{e}\left[\mathcal{C}_{en,ex}\right]_{2k} = \sum_{j=0}^k \left(-1\right)^{j+1} 2^{(j+1)} \binom{k}{j+1} R_{ex}^{\left(k-j\right)} \bar{\phi_{ex}}^{j+1}
\end{equation}
Because of the equal sharing hypothesis, Equation~\ref{eqn:moments-of-boltzmann-operator} for ionization collisions differs
from that for excitation collisions:
\begin{equation}
\begin{split}
	&M^{e}\left[\mathcal{C}_{en,iz}\right]_{2k} = \dfrac{1-2^{k-1}}{2^{k-1}}R_{iz}^{\left(k+1\right)} + \dots\\
    &\dots \dfrac{1}{2^{k-1}}\sum_{j=0}^k \left(-1\right)^{j+1} 2^{(j+1)} \binom{k}{j+1} R_{iz}^{\left(k-j\right)} \bar{\phi_{iz}}^{j+1},
    \end{split}
\end{equation}
with $M^{e}\left[\mathcal{C}_{en,iz}\right]_{0} = R^{(1)}_{iz}$. The odd-order moments of the ionization collision operator are also uniformly
zero because we neglected momentum exchange contributions of inelastic collisions as compared to that of elastic collisions.
The first seven parallel moments for the ionization collision operator are:
\begin{equation}
M^{e}\left[\mathcal{C}_{en,iz}\right]
=	 \begin{bmatrix}
		R^{(1)}\\
		0\\
		-2\bar{\phi} R^{(1)}\\
		0 \\
        -\dfrac{1}{2} R^{(3)} - 2\bar{\phi} R^{(2)} + 2\bar{\phi}^2 R^{(1)}\\ 
		0\\
        -\dfrac{3}{4} R^{(4)}-\dfrac{3}{2}\bar{\phi} R^{(3)} + 3\bar{\phi}^2 R^{(2)} - 2\bar{\phi}^3 R^{(1)}
	\end{bmatrix}.
\end{equation}

\subsubsection{System of equations}
Using Equations~\ref{eqn:inelastic-moments-multinomial} and the approximated VDF, 
we now have a system of ODE
for the moments of the species of interest, $M_{ijk}$,
and moments of the collision operator, $M^{e}\left[\mathcal{C}_{en}\right]$.
For the first seven moments, it is given by:
\begin{equation}
\dfrac{\diff}{\diff t} 
\begin{bmatrix}
    M_0 \\
    M_{2} + 2 M_0 T\\
    M_{4} 
    +4 M_2\bar{T} 
    +8 M_0\bar{T}^2 \\ 
   M_6 + 6 M_4 \bar{T} + 24 M_2 \bar{T}^2 + 48 M_0 \bar{T}^3 
    \\ 
\end{bmatrix}
= 
\begin{bmatrix}
M^{e}\left[\mathcal{C}_{en}\right]_0\\
M^{e}\left[\mathcal{C}_{en}\right]_2\\
M^{e}\left[\mathcal{C}_{en}\right]_4\\
M^{e}\left[\mathcal{C}_{en}\right]_6
\end{bmatrix},
\end{equation}
where $\bar{T} = k_B T / m$.
This system of ODE is closed with the relationship $T = M_2/M_0-M_1^2/M_0^2$.
Assuming that the reaction integrals are held constant over the small integration step,
this system can be recast as a system of first-order ODEs in the even-order moments,
and an explicit solution may be found for all moments.  
In practice, we solve this system using the forward Euler method over a half time step
in the corresponding Strang splitting step.

\subsection{Evaluation of collision integrals}

Particle scattering is inherently three-dimensional, and, therefore, remains challenging to implement 
within a quadrature-based moment method framework.
Berger~\textit{et al.}\cite{Berger2025} limit their application of HyQMOM to collisions that 
feature purely back-scattering (\textit{e.g.}, charge-exchange collisions) or that are similar to BGK collisions
so that the 1D3V problem is simplified to a 1D1V one.
However, because we assumed equilibrium in the transverse direction 
and that the VDF is a sum
of individual kernel functions in the parallel direction, 
reaction integrals may be simplified.
The exact shape of the reaction integral depends on the quadrature method chosen. 
We list in the next sections results for the reaction integrals in both 1D1V and 1D3V. 
In all cases, the collision cross section data are sourced from the LXCat Project.\cite{Pancheshnyi2012,Pitchford2017}

 \subsubsection{Elastic collisions}

The reaction rate may be computed with Equation~\ref{eqn:reaction-rate-generic}, which is a 6D integral.
With the approximate VDF considered (Equation~\ref{eqn:simplifying-transverse-equilibrium})
we have:
\begin{equation}
\begin{split}
	R_{ij} \approx 
  &  \sum_{\alpha_i}\sum_{\alpha_j}
    \omega_{\alpha_i}
    \omega_{\alpha_j}
    \left( \dfrac{m_i}{2 k_B T_{\perp,i}} \right)
    \left( \dfrac{m_j}{2 k_B T_{\perp,j}} \right)
    \cdot\\
    &\iint_{\mathds{R}^3\times\mathds{R}^3} \left|\mathbf{v}_i - \mathbf{v}_j\right| 
    Q\left(\left|\mathbf{v}_i - \mathbf{v}_j\right|\right)\\
    &\exp\left(-\dfrac{m_i}{2 k_B T_{\perp,i}} \left(v^2_{y,i} + v^2_{z,i}\right)\right)
    \exp\left(-\dfrac{m_j}{2 k_B T_{\perp,j}} \left(v^2_{y,j} + v^2_{z,j}\right)\right)\\
    &K_i\left(v_{\alpha_i};\mathbf{v}_i\right)
    K_j\left(v_{\alpha_j};\mathbf{v}_j\right)
	\diff^3 \mathbf{v}_i \diff^3\mathbf{v}_j,
    \end{split}
\end{equation}
where $\omega_{\alpha_i}$ and $v_{\alpha_i}$ are the weights and abscissas for the species $i$ and QBMM under consideration,
and $K_i\left(v_{\alpha_i}\right)$ is the QBMM kernel used to represent the VDF of species $i$.
For simplicity, we will consider in this work that both species $i$ and $j$ are represented with the same QBMM and number of nodes. 

For 1D1V cases (\textit{i.e.}, for cases where the transverse exponential component collapses
to a Dirac delta function centered at zero velocity),
and 
assuming that the VDF for both species is approximated using HyQMOM or GQMOM,
the reaction rate reduces to a double sum because the kernels are Dirac delta functions:
\begin{equation}
    R_{ij} \approx 
    \sum_{\alpha_i}\sum_{\alpha_j}
    \omega_{\alpha_i}
    \omega_{\alpha_j}
    \left|v_{\alpha_i} - v_{\alpha_j}\right| 
    Q\left(\left|v_{\alpha_i} - v_{\alpha_j}\right|\right).
		\label{eqn:approximation-elastic-en-rate-hyqmom}
\end{equation}

The 1D3V case is more challenging, and it is not possible to further reduce the 6D integral when
the QBMM considered is either HyQMOM and GQMOM.
However, the 6D integral may be reduced to a 1D integral in the EQMOM case of Gaussian kernels. 
For Gaussian kernels, the integral is given by:
\begin{equation}
\begin{split}
R_{ij} \approx 
  &  \sum_{\alpha_i}\sum_{\alpha_j}
    \dfrac{
    \omega_{\alpha_i}
    \omega_{\alpha_j}
    }{
    \sqrt{\left(2\pi \sigma_i^2\right)\left(2\pi \sigma_j^2\right)}
    }
    \left( \dfrac{m_i}{2 k_B T_{\perp,i}} \right)
    \left( \dfrac{m_j}{2 k_B T_{\perp,j}} \right)
    \\
    &\cdot
    \iint_{\mathds{R}^3\times\mathds{R}^3} \left|\mathbf{v}_i - \mathbf{v}_j\right| 
    Q\left(\left|\mathbf{v}_i - \mathbf{v}_j\right|\right)\\
    &
    \exp\left(-\dfrac{m_i}{2 k_B T_{\perp,i}} \left(v^2_{y,i} + v^2_{z,i}\right)\right)
    \exp\left(-\dfrac{m_j}{2 k_B T_{\perp,j}} \left(v^2_{y,j} + v^2_{z,j}\right)\right)\\
    &\exp\left(-\dfrac{1}{2}\dfrac{\left(v_i-v_{\alpha_i}\right)^2}{\sigma_i^2}\right)
    \exp\left(-\dfrac{1}{2}\dfrac{\left(v_j-v_{\alpha_j}\right)^2}{\sigma_j^2}\right)
	\diff^3\mathbf{v}_i \diff^3\mathbf{v}_j.
		\label{eqn:approximation-elastic-en-rate-eqmom}
        \end{split}
\end{equation}
This is the reaction rate for interacting, drifting bi-Maxwellian distributions.
Each bi-Maxwellian distribution has a transverse temperature $T_\perp$ and an equivalent
parallel temperature, $T_\parallel = m \sigma^2 / k_B$.
Closed-form expressions for this particular integral are given in Ref.~\onlinecite{Xie2023} 
and repeated in the Appendix.
The single, one-dimensional, integral in energy space may be evaluated with a simple trapezoid method or an adaptive quadrature 
(\textit{e.g.}, Gauss-Kronrod).

\subsubsection{Inelastic collisions}

\begin{figure}[ht!]
     \centering
     \includegraphics[scale=0.75]{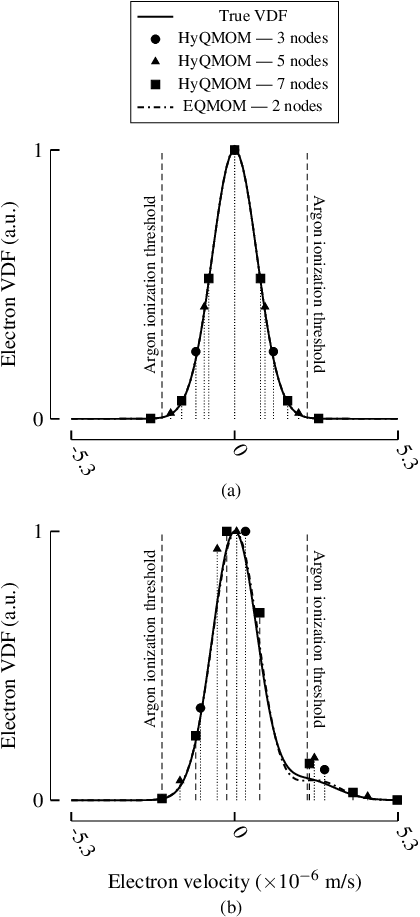}
     \caption{Example electron VDF and HyQMOM and EQMOM representation. (a) Maxwellian VDF with zero-drift and $T_e = 3$~eV.
     (b) Superposition of two Maxwellian VDFs with $T_e = 3$~eV and zero-drift, and $T_e = 5$~eV and drift velocity of
     2.3$\times10^6$ m/s. EQMOM exactly recovers the VDF in (a) and the two curves are indistinguishable.}
     \label{fig:maxwellian-vdf-quadratures}
 \end{figure}

 Using Equation~\ref{eqn:reaction-rate-inelastic}, the inelastic reaction integrals in 1D1V (\textit{i.e.}, $f_y = f_z = \delta(v)$) are given by:
 \begin{equation}
 	R^{(k)} \approx n_n 
	\sum_{\alpha}
    \omega_\alpha
    \int_{\mathds{R}}
	\left|\mathbf{v}_e\right|^{2k-1}
    K\left(v_\alpha\right) 
	Q^T\left(\left|\mathbf{v}_e\right|\right)
	\diff\mathbf{v}_e,
 \end{equation}
 where $K$ is the kernel function considered.
 For Dirac kernel functions (\textit{e.g.}, HyQMOM or GQMOM representation of the VDF), the reaction integral 
 is greatly simplified and, much like Equation~\ref{eqn:approximation-elastic-en-rate-hyqmom}, reduces to a sum:
 \begin{equation}
  	R^{(k)} \approx n_n 
	\sum_{\alpha}
    \omega_\alpha
	\left|\mathbf{v_\alpha}_e\right|^{2k-1}
	Q^T\left(\left|\mathbf{v_\alpha}_e\right|\right).
 \end{equation}
 While this form of the reaction integral is attractive due to its simplicity, it is challenging to obtain
 accurate values for threshold processes such as ionization or excitation
 with a low number of nodes.
 Because $T_e \approx$1--5~eV for the plasmas considered, a HyQMOM representation of the VDF might result in abscissas
 that are well below the threshold energy ($\approx$10--20~eV). This results in a reaction integral that is uniformly zero.
 We illustrate this issue in Figure~\ref{fig:maxwellian-vdf-quadratures}, which displays both a Maxwellian (a) and non-Maxwellian (b) 1D1V distribution
 function of electrons along with its HyQMOM and EQMOM representations and
 the threshold electron velocity required for a non-zero ionization collision cross section in argon.
 In the Maxwellian case, HyQMOM with 3--5 nodes results in abscissas placed at velocities lower than that of the 
 ionization threshold velocity, which results
 in a uniformly zero reaction integral.
 By contrast, the EQMOM representation of the VDF coincides exactly with the true VDF 
 because it uses Gaussian kernels.
 In the non-Maxwellian case, at least one node is placed above the ionization velocity threshold and the reaction integral
 is non-zero. However, there is no guarantee that HyQMOM with a few nodes can provide a non-zero reaction integral. 
 The EQMOM representation yields a close representation of the true VDF
 for $v\leq 0$~m/s, and is able to capture the tail bump at $v\approx 2.3\times10^6$~m/s.
 In general, using the EQMOM representation should result in a numerical evaluation of collision integrals that is within a 
 few percent of the true reaction integrals.

 For Gaussian kernels (\textit{i.e.}, EQMOM representation of the VDF) in 1D1V,
 a Gauss-Hermite \textit{secondary} quadrature may be used to compute
 Equation~\ref{eqn:reaction-rate-inelastic}, 
 as suggested by Yuan~\textit{et al.}\cite{Yuan2012}:
    \begin{align}
		\label{eqn:approximation-inelastic-en-rate-eqmom}
        R^{(k)} &\approx n_n 
        \sum_{\alpha} \dfrac{\omega_\alpha}{\sqrt{\pi}}
        \int_{\mathds{R}}
        \left|z\right|^{2k-1}
        \exp\left(-z^2\right)
        Q^T\left(\left|z\right|\right)
        \diff z\\
        &\approx
        n_n
        \sum_{\alpha} 
        \dfrac{\omega_\alpha}{\sqrt{\pi}}
        \sum_{\beta}
        \omega_{\alpha\beta}
        \left|z_{\alpha\beta}\right|^{2k-1}
        Q^T\left(\left|z_{\alpha\beta}\right|\right)
    \end{align}
  where $z = \left(v-v_\alpha\right)/\sigma\sqrt{2}$, $\beta$ is the index for the Gauss-Hermite nodes,
  and $\omega_{\alpha\beta}$ is the weight of the Gauss-Hermite quadrature.
  This Gauss-Hermite quadrature is exact if the integrand ($
        \left|z\right|^{2k-1}
        Q^T\left(\left|z\right|\right)
  $) is a polynomial function.
  In general, however, the cross section cannot be easily approximated as a polynomial
  function of the electron velocity, and a large number of nodes is required to
  accurately compute the reaction rate with a Gauss-Hermite quadrature.
  Other secondary quadratures (\textit{e.g.}, Gauss-Kronrod) may be used but require the inclusion of the exponential
  term in the integrand ($\exp\left(-z^2\right)$). 

We resort again to EQMOM for the more general 1D3V case: 
\begin{equation}
\begin{split}
 	R^{(k)} 
    &\approx n_n 
    \sum_{\alpha}
    \dfrac{
    \omega_{\alpha}
    }{
    \sqrt{\left(2\pi \sigma^2\right)}
    }
    \left( \dfrac{m}{2 k_B T_{\perp}} \right)
    \cdot
    \int_{\mathds{R}^3} \left|\mathbf{v}_e\right| 
    Q\left(\left|\mathbf{v}_e\right|\right)\\
    &
    \cdot\exp\left(-\dfrac{m_i}{2 k_B T_{\perp}} \left(v^2_{y} + v^2_{z}\right)\right)
    \exp\left(-\dfrac{1}{2}\dfrac{\left(v_e-v_{\alpha}\right)^2}{\sigma^2}\right)
    \diff^3\mathbf{v}_e.    
\end{split}
\end{equation}
This 3D integral may also be reduced to a single dimension.

\subsubsection{QBMM limitations}

    \begin{figure*}[t]
        \centering
        \includegraphics[scale=1.1]{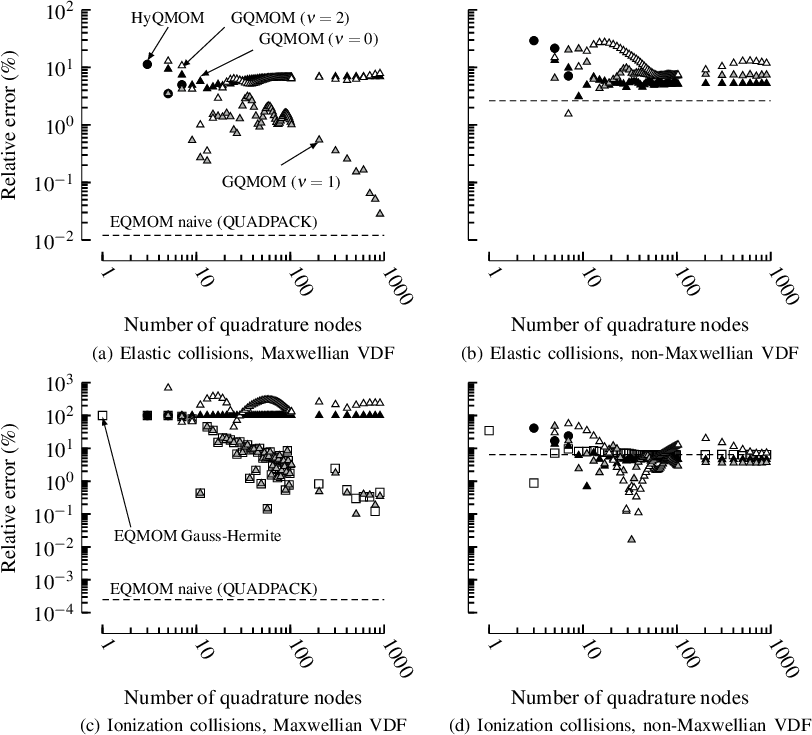}
        \caption{Relative error for the computation of the collision reaction rate for (a,b) elastic and 
        (c,d) ionization collisions of an electrons impacting argon neutral atoms. Electrons follow either a (a,c) Maxwellian VDF 
        or a (b,d) non-Maxwellian VDF. Both VDFs are shown in Figure~\ref{fig:maxwellian-vdf-quadratures}.}
        \label{fig:reaction-integral-error}
    \end{figure*}

  We now illustrate the different approaches 
  to compute the reaction integrals 
  (HyQMOM, GQMOM, EQMOM both with and without secondary quadrature)
  for two different 1D1V examples.
  In our first two examples, electrons are considered to follow the VDFs shown 
  in Figure~\ref{fig:maxwellian-vdf-quadratures}(a) and (b).
  The electrons collide with argon atoms that are kept at a temperature of 300~K.
  For simplicity, the VDF in the transverse direction is assumed for these examples to be a Dirac delta
  function centered at zero, so that only a single velocity direction is considered in the reaction rate integrals: $\textbf{v}_e = v_e \in \mathds{R}$. 
  As a result, the 6D and 3D integrals that appear in the reaction rate calculations are simplified
  to a 2D and 1D integral, respectively.
  The elastic and ionization cross section data are obtained from the Hayashi database.\cite{LXCat-Hayashi,Hayashi2003}

  For the two example VDFs considered, 
  the relative error 
  for both the elastic and ionization reaction integral of order $1$ (\textit{i.e.}, $R^{(1)}$)
  is shown in Figure~\ref{fig:reaction-integral-error} 
  as a function of the number of quadrature nodes. 
  We compute the ``true'' reaction integral with either:
  \begin{itemize}
      \item a quadrature from QUADPACK\cite{Piessens2012} for the double integral that appears in elastic collisions, as implemented in SciPy,\cite{SciPy2020} or
      \item a simple trapezoid rule for the single integral that appears in inelastic collisions.
  \end{itemize}
  In both cases, the neutral and electron velocities are limited to $\pm 15$ times that of the thermal velocity. 

  HyQMOM is able to provide numerical estimates of the elastic reaction within 10\% of the ``true''
  reaction rate (Figures~\ref{fig:reaction-integral-error} (a--b)). 
  However, as discussed in the previous section, because only a few nodes are above the ionization
  velocity threshold, at least seven nodes are required to have a non-zero ionization reaction rate.
  Even with seven nodes, the corresponding relative error for HyQMOM remains high ($>90$\%
  and $>40$\% for the Maxwellian and non-Maxwellian VDF, respectively) for the ionization rate (Figures~\ref{fig:reaction-integral-error} (c--d)).
  
  GQMOM introduces many more quadrature nodes (up to 1,000 in this study) to reduce the integration error.
  However, the behavior of GQMOM varies significantly depending on the value of the free parameter $\nu$:
  for a Maxwellian VDF, $\nu=1$ is ideal as it is able to recover the Gauss-Hermite quadrature (see, \textit{e.g.},
  Figure~\ref{fig:reaction-integral-error}(c)). 
  For more general VDFs, however, $\nu=1$ can still result in significant error in the estimation of the reaction rates (see, \textit{e.g.}, Figures~\ref{fig:reaction-integral-error}(b) and (d)).
  It remains unclear how the shape of the integrand and the total number of nodes affect the optimal value 
  of $\nu$ that minimizes the relative error.

  The na{\"i}ve integration of Equations~\ref{eqn:approximation-elastic-en-rate-eqmom} and \ref{eqn:approximation-inelastic-en-rate-eqmom}
  with QUADPACK results in the smallest error in the reaction rates in most cases
  (see, \textit{e.g.}, Figures~\ref{fig:reaction-integral-error}(a--c)), although
  GQMOM outperforms QUADPACK within a certain range of number of nodes. 
  While the na{\"i}ve approach to obtain the elastic collision rate seems attractive, it
  requires the costly numerical evaluation of a double integral for 0D1V and 1D1V problems 
  where the transverse VDF is assumed to be a Dirac delta function.
  In the more general 0D3V or 1D3V case, however, the 6D and 3D integrals for both the elastic
  and inelastic reaction rates may be reduced to a single 1D integral, which may be 
  evaluated efficiently.

\subsubsection{Reaction rate validation for threshold reactions}

We now compare our ionization rate for argon to that obtained with the Grad-9M method of Alvarez-Laguna~\textit{et al.}\cite{Alvarez2022}
for multiple non-drifting electron VDF, with three velocity directions.
We use the dataset from Phelps\cite{LXCAT-Phelps} for the ionization collision cross section. 
	  Prior to conducting the reaction integrals, the fourth moment of the VDF is moderately perturbed from thermal equilibrium such that $-0.1\leq \Delta \leq 0.1$, where
      $\Delta$ is the excess kurtosis of the VDF. 
      The excess kurtosis is defined here 
    as the fourth-order contracted moment normalized by the
    second-order moment of the equilibrium VDF:\cite{Alvarez2022,Wu2026} 
    \begin{equation}
        \Delta = \dfrac{\dfrac{1}{2}\displaystyle\int_{\mathds{R}^3} \left(\mathbf{v}-\bar{\mathbf{v}}\right)^4 f\diff^3\mathbf{v}}{\dfrac{15}{2} M_0 \left(k_B T/m\right)^2}-1,
    \end{equation}
    where $\bar{\mathbf{v}}$ is the drift velocity.
    For the Grad-9M moment method, a non-zero value of $\Delta$ results in an isotropic
    perturbation of the VDF from thermal equilibrium.
    Using the definition of our approximate VDF ($\tilde{f}$, Equation~\ref{eqn:simplifying-transverse-equilibrium}), $\Delta$ is given by:
    \begin{equation}
    \Delta = \dfrac{M_{400} + 12 M_0 \left(k_B T /m\right)^2}{15 M_0 \left(k_B T / m\right)^2} -1, 
    \end{equation}
    such that $\Delta=0$ corresponds
    to a Maxwellian VDF in all directions.
    A non-zero value of $\Delta$ results in a perturbation of the fourth order moment 
    of the parallel VDF because of the chosen approximate VDF. 
    For this particular definition of the excess kurtosis, $\Delta$ must be greater than -0.2  to ensure that $M_4$ remains positive for our VDF.  
    Additionally, $\Delta$ must be negative to
    ensure that $M_4/M_0 \leq 3 M_2^2/M_0$ in the two-node EQMOM algorithm in the case where $M_1 = M_3 = 0$.\cite{Chalons2017}
    To remove the latter constraint, we use a three-node EQMOM algorithm for positive values of $\Delta$. 
    The fourth and sixth-order moment ($M_{400}$ and $M_{600}$) are then computed using the Grad-9M definition of the VDF. 

\begin{figure}[ht!]
    \centering
    \includegraphics[scale=1.0]{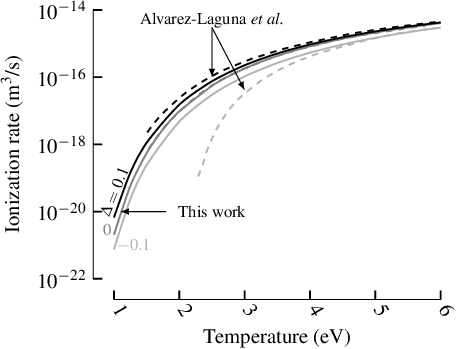}
    \caption{Comparison of the ionization rate as obtained with QBMM and with Grad 9M method as a function of bulk electron temperature, for multiple excess kurtosis values.}
    \label{fig:ionization-rate-comparison}
    \end{figure}

    The ionization rate is shown in Figure~\ref{fig:ionization-rate-comparison} for three values of $\Delta$.
    Exact agreement with Grad's method is obtained for the thermal equilibrium case.
    Although an exact match is not expected as the excess kurtosis is a parameter that controls
    the shape of the VDF in \textit{all} directions in Grad's method 
    while it controls only the parallel direction in this work, 
    good agreement is obtained between the two methods in most cases.
    The expected trends of higher ionization rate with increasing electron temperature
    and with increasing $\Delta$ are captured with QBMM. For the leptokurtic case ($\Delta >0$), additional
    high-energy electrons are present in the tail of the VDF, which contributes to increasing the reaction rate.
    Conversely, for the mesokurtic case ($\Delta < 0$) the tails of the VDF are depleted from their high-energy electrons,
    and, consequently, the ionization rate is lower.
    
The discrepancy observed at low temperatures and negative values of $\Delta$ 
is due to the fact that the VDF from the Grad-9M method may become negative.
In those instances, the negative values of the VDF are clamped to a zero value,\cite{Alvarez2022} which yields a fast
decay towards zero ionization rate at lower temperatures.
For a non-drifting distribution, the isotropic multiplication term in Grad's method is:
\begin{equation}
    g = 1 + \Delta \left(\dfrac{15}{8}-\dfrac{5}{2} \bar{E} + \dfrac{1}{2}\bar{E}^2\right),
\end{equation}
where $\bar{E} = m \left(v^2_x+v^2_y+v^2_z\right)/(2 q T_{eV}) = E/T_{eV}$ is the normalized particle energy.
The ionization cross section is non-zero if ${E} > E_{iz} \Leftrightarrow \bar{E} > E_{iz}/T_{eV}$,
where $E_{iz}$ is the ionization energy.
For a given excess kurtosis that is negative, 
$g>0$ (and, therefore, the VDF is positive) for $\bar{E}< \frac{5}{2} + \sqrt{\frac{5}{2} - \frac{2}{\Delta}}$.
This constraint yields a minimum temperature such that the reaction integral is non-zero using Grad's approach: 
\begin{equation}
    T_{eV,\textrm{min}} = \dfrac{\epsilon_{iz}}{\dfrac{5}{2} + \sqrt{\dfrac{5}{2} - \dfrac{2}{\Delta}}}.
\end{equation}
For bulk electron temperatures below that threshold, the ionization rate
is exactly zero, as illustrated in Figure~\ref{fig:ionization-rate-comparison}:
for $\Delta = -0.1$, $T_{eV,\textrm{min}} \approx 2.2$~eV. 
In practice, however, significant error can still arise in the computation of the ionization rate 
even if $T_{eV} > T_{eV,\textrm{min}}$ as a large
portion of the VDF in the Grad-9M method might still be exactly zero. 
For example, there is a large discrepancy between the reaction rates obtained with QBMM and Grad 9M method
for $\Delta = -0.1$ and $T_{eV,\textrm{min}} < T_{eV} < 4$~eV.
This numerical example illustrates one of the advantages of QBMM over Grad's method: 
by construction, the VDF obtained with QBMM cannot
be negative, which, in turn, allows us to compute the ionization rate even for 
electron temperatures that are below $T_{eV,\textrm{min}}$.

%%%%%%%%%%%%%%%%%%%%%%%%%%%%%%%%%%%%%%%%%%%%%%%%%%%%%%%%%%%%%%%%%%%%%%%%%%%%%%%%%%%%%%%%%%%%%%%%%%%%%%%%%%%%%%%%%%
\section{Results and discussion}

    The implementation of the collision models and moment method is now validated on four test cases of increasing complexity.
	The electron-electron, electron-neutral, and RF test cases are outlined in Refs.~\onlinecite{Alvarez2022,Turner2013}.
    When possible, we compare our results to that obtained by the original authors, regardless of the numerical algorithm
    chosen by the original authors (\textit{i.e.}, whether a moment method was used).

\subsection{Coulomb collisions}

\subsubsection{Electron-ion}
    
We first consider the temperature relaxation between electrons and protons within a zero-dimensional cell (no moment transport).
	 Only electron-ion Coulomb collisions are considered for this example: self-collisions are neglected. 
The initial VDF for both species is either a non-drifting Maxwellian VDF, or a drifting non-Maxwellian VDF given by:
	 \begin{equation}
	 	f_s\left(v\right) = A \left(1+ \dfrac{1}{2}\cos\left(\dfrac{\pi}{v_{th,s}} v\right)\right)
	 	\exp\left(-\dfrac{\left(v-u_s\right)^2}{v_{th,s}^2}\right),
	 \end{equation}
	 where $u_s$ and $v_{th,s}$ are the bulk and thermal velocities of species $s$, respectively, and $A$ is a normalization constant.
	 For the non-Maxwellian case, $u_s = v_{th,s}/2$.
     In both cases, the initial electron and ion temperature is $40$~eV and $80$~eV, respectively.
	 
\begin{figure}[ht!]
	\centering
    \includegraphics[scale=0.85]{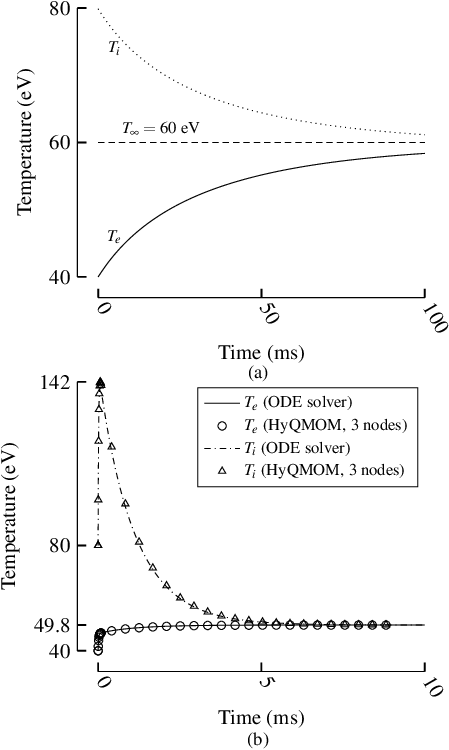}
	\caption{Electron and proton temperature for (a) a non-drifting, Maxwellian VDF, and (b) a drifting, non-Maxwellian VDF. Results of a direct ODE solver applied to Equations~\ref{eqn:coulomb-collisions-electron} and~\ref{eqn:coulomb-collisions-ion} are also shown in (b).} 
	\label{fig:coulomb-collisions}
	\end{figure}

Figure~\ref{fig:coulomb-collisions} shows the electron and ion temperature as a function of time
as computed with QBMM for the (a) Maxwellian and (b) non-Maxwellian initial VDF. 
Because we considered that $n_e = n_i$, the charged particle temperature simply relaxes 
towards $T_\infty = 1/2\left(T_{e0} + T_{i0}\right) = 60$~eV in the Maxwellian case.\cite{Nanbu1997}
  As shown in Figure~\ref{fig:coulomb-collisions}(b), despite the initial non-Maxwellian nature of the test VDFs, 
  HyQMOM with three nodes (\textit{i.e.}, five moments) is able to accurately capture 
  the relaxation of both the electron and proton VDF towards a Maxwellian distribution with a temperature
	 of $\approx$50~eV.
     This temperature differs from the Maxwellian case because the initial VDF is both drifting and non-Maxwellian.

     \begin{figure}[ht!]
	 	\centering
	 	\includegraphics[scale=0.8]{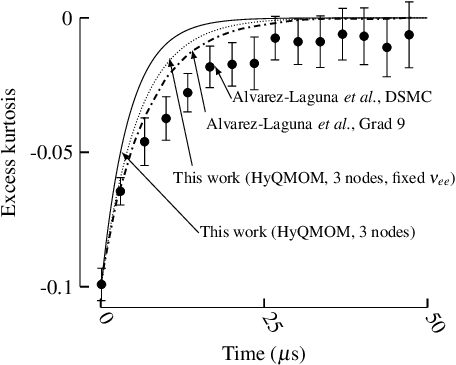}
	 	\caption{Evolution of the excess kurtosis for multiple moment models and DSMC for electron-electron collisions
	 		with $n_e = 10^{17}$~m\textsuperscript{-3}, $\mathbf{u}_e$ = 0~m/s, $T_e=5$~eV, and $\Delta(t=0)=-0.1$.}
	 	\label{fig:electron-electron-relaxation}
	 \end{figure}

        	 \begin{figure*}
	 	\centering
	\includegraphics[scale=0.90]{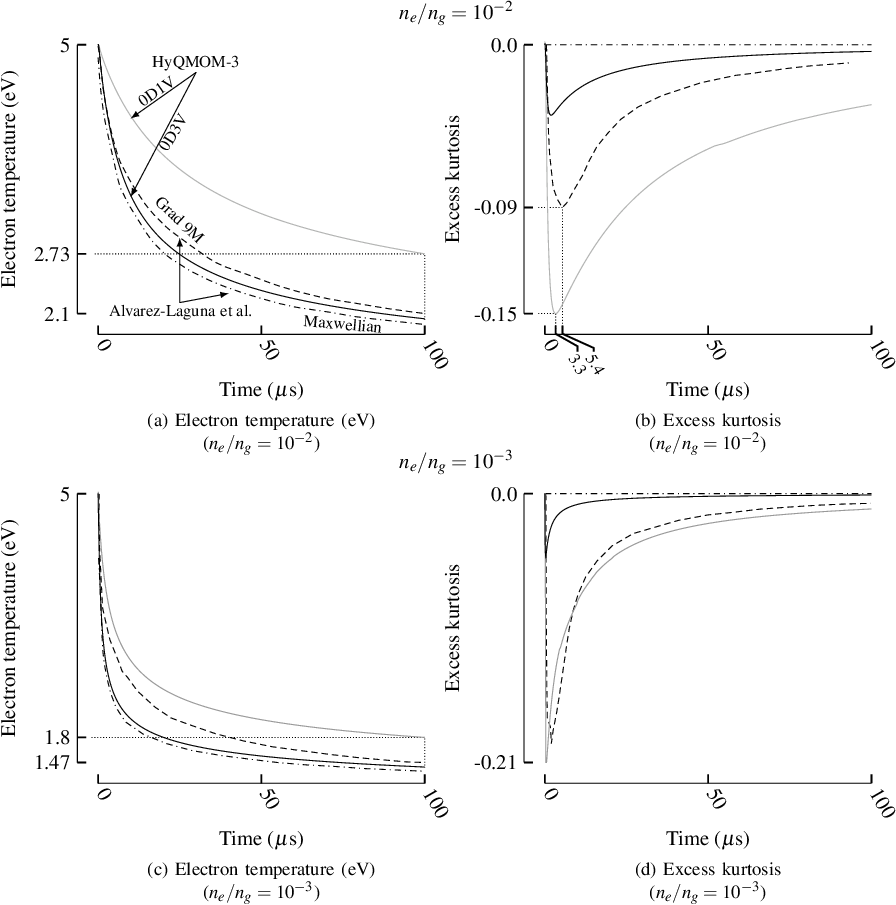}
	\caption{Evolution of the electron temperature and excess kurtosis for multiple moment models for electron-argon collisions
		with $n_e = 10^{17}$~m\textsuperscript{-3}, $\mathbf{u}_e$ = 0~m/s, $T_e=5$~eV, $\Delta(t=0)=0$, $\mathbf{u}_g$ = 0~m/s, and $T_g = 300$~K.
		(a--b) $n_g = 10^{19}$~m\textsuperscript{-3}, (c--d)  $n_g = 10^{20}$~m\textsuperscript{-3}.
		}
	\label{fig:electron-argon-collisions}
	\end{figure*}
	 	  
	 \subsubsection{Electron-electron}
     
     Consider now the relaxation of a perturbed electron VDF towards a Maxwellian.\cite{Alvarez2022} 
     Only electron-electron Coulomb collisions occur within a zero-dimensional cell.
	 The initial electron population has a density, drift velocity, and temperature of $10^{17}$~m\textsuperscript{-3}, 0~m/s, and 5~eV, 
     respectively, and the fourth moment of the VDF is initially perturbed from thermal equilibrium such that $\Delta = -0.1$. 

	 Figure~\ref{fig:electron-electron-relaxation} shows the excess kurtosis as a function of time.
	 Good agreement is obtained with the results of Alvarez-Laguna~\textit{et al.}\cite{Alvarez2022} 
  The difference in relaxation rate between our approach and that of the Grad-9M method 
  is attributed to a difference in implementation of
     the collision operator and in the choice of the approximate VDF.
     Closer agreement with the results of Alvarez-Laguna~\textit{et al.} can be obtained by setting 
     the collision frequency to a constant value of $\nu_{ee}\approx 1.8\times 10^5$s\textsuperscript{-1}.
  In all cases, successive electron-electron collisions restore thermal equilibrium (\textit{i.e.}, $\Delta \rightarrow 0$ as $t\rightarrow +\infty$).

	 \subsection{Collisions with neutral background}
    
      \begin{figure*}[t]
	\includegraphics[scale=0.90]{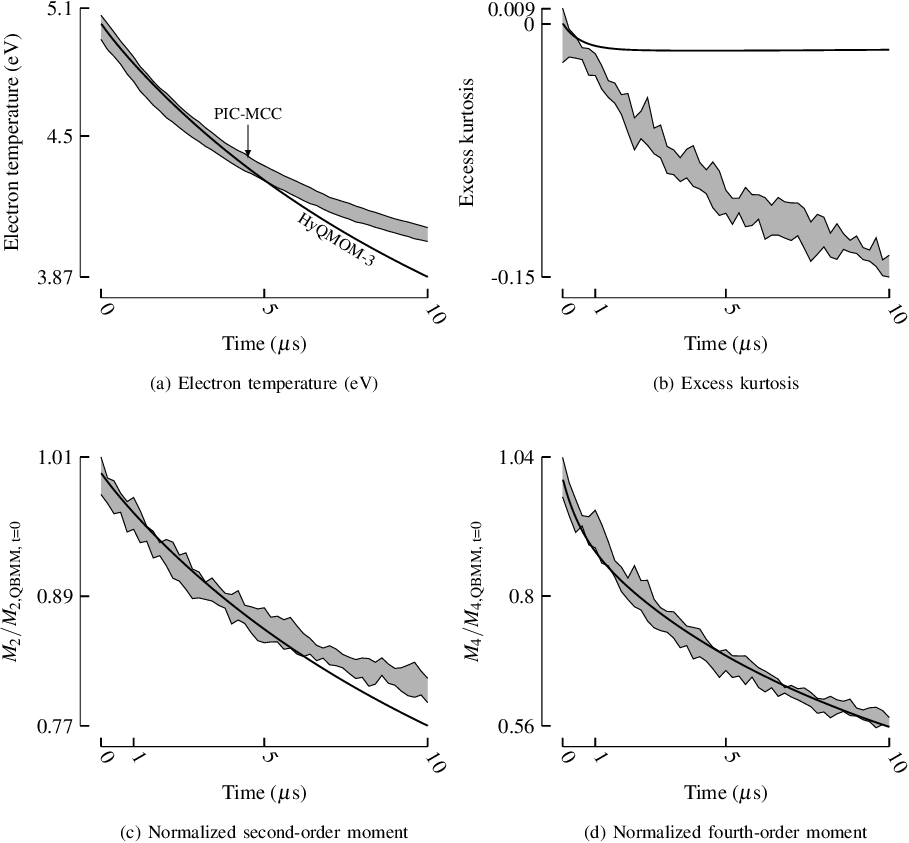}
    \caption{
    (a) Electron temperature, (b) excess kurtosis, and normalized (c) second-order and (d) fourth-order moments, $M_2$ and $M_4$.  
    The second and fourth-order moments are normalized by their respective value at $t=0$.
    The data is for the electron-argon collisions test case, without
    elastic and ionization collisions.}
    \label{fig:electron-argon-collisions-no-ee-m2-m4}
    \end{figure*}
     
	We next consider the evolution of the electron VDF due to elastic (electron-electron and electron-neutral) and 
	inelastic (ionization and excitation) collisions with argon atoms.
	The argon gas is at thermal equilibrium 
    with zero bulk velocity and a temperature of 300~K, and with a density $n_g$ such that the ionization fraction, $n_e/n_g$, is either $10^{-2}$ or $10^{-3}$.
	Electrons are initially at equilibrium with a density of $10^{17}$~m\textsuperscript{-3}, zero bulk velocity, and a 
	temperature of 5~eV.
	We maintain the electron density constant: ionization collisions do not contribute to an increase in electron density.

	Figure~\ref{fig:electron-argon-collisions} shows the electron temperature and excess kurtosis as a function of time, for two
	different ionization fractions.
	QBMM are able to qualitatively reproduce the temperature decay, 
	as well as the depletion of the tail of initial Maxwellian distribution 
	(\textit{i.e.}, increase in $\left|\Delta\right|$)
	 and the
	thermalization as $t\rightarrow +\infty$ (\textit{i.e.},  $\left|\Delta\right|\rightarrow 0$).
    An exact match between our results and that of Alvarez-Laguna \textit{et al.}\cite{Alvarez2022} 
    is not expected because (i) we assume equilibrium in both transverse directions and
    (ii) the implementation of the electron-neutral and electron-electron elastic collision operators 
    is different: Alvarez-Laguna~\textit{et al.} use a Boltzmann operator for elastic collisions while we use a BGK model for simplicity, 
    and the shape of the integrand for the threshold processes differs from ours.

   \begin{figure*}
	 	\centering
	\includegraphics[scale=0.90]{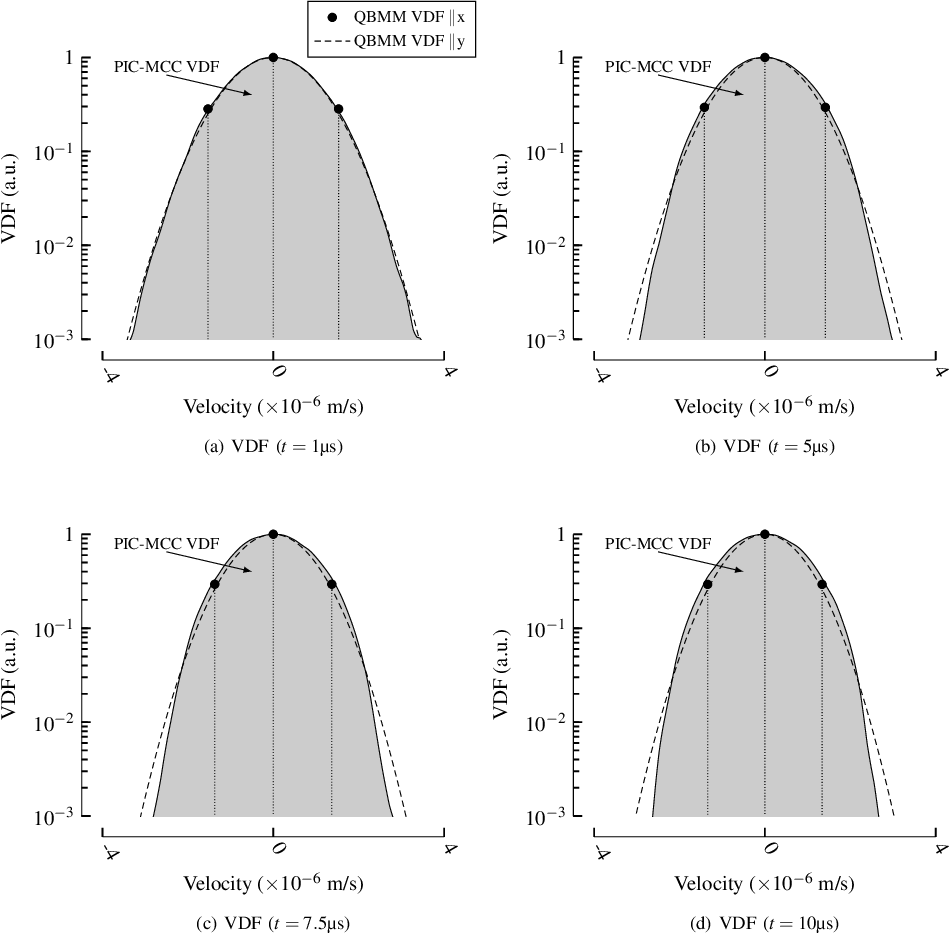}
	\caption{Isotropic VDF for the excitation collision test case in argon, at four different times (a) \SI{1}{\micro\second}, 
    (b) \SI{5}{\micro\second}, (c) \SI{7.5}{\micro\second}, and (d) \SI{10}{\micro\second}. 
    The VDF is normalized by its largest value.
		}
	\label{fig:electron-argon-collisions-no-ee}
	\end{figure*}

    To what extent are QBMM able to model further deviations from equilibrium?
    We now explore one of the limitations of our approach and choice of BGK model for electron-neutral collisions and
    remove the electron-electron collisions that thermalize the VDF.
    For simplicity, we also remove ionization collisions and consider only elastic electron-neutral collisions and excitation collisions
    and consider only the test case for which $n_e/n_g = 10^{-2}$.
    The electron temperature, excess kurtosis, as well as the non-constant, even-order moments that would be transported with HyQMOM with three nodes, $M_2$ and $M_4$, 
    are shown in Figure~\ref{fig:electron-argon-collisions-no-ee-m2-m4}. 
    The $M_2$ and $M_4$ moments are normalized by the corresponding moment of the VDF at $t=0$.
    The VDF is shown for $t\leq 10$\unit{\micro\second} in Figure~\ref{fig:electron-argon-collisions-no-ee}.
    Our results are compared to that of a modified version of ``mini-PIC,''\cite{Powis2024} a PIC-MCC implementation. For this particular test case, the PIC solver is run
    with 16,384 particles and a time step $\Delta t = 10^{-8}$s.
    The minimum and maximum of each quantity for five PIC-MCC runs are displayed on 
    Figure~\ref{fig:electron-argon-collisions-no-ee-m2-m4}.
    The PIC-MCC VDF is computed using a Kernel Distribution Estimate (KDE) of the instantaneous particle velocity.
    A Gaussian kernel is chosen for the KDE, with a cross-validated bandwidth. 
    The cross-validation process consists of a K-fold method with five folds and a bandwidth
    that is in the range of $\left[0.1\bar{h},10\bar{h}\right]$, where $\bar{h} = 1.06 \sigma/N^{0.2}$ is an
    initial bandwidth estimate, $\sigma$ is the PIC VDF standard deviation, and $N$ is the total number of velocity samples.
    Because the PIC-MCC VDF is isotropic, only its $x$-direction component is displayed. 
    
    Good agreement between QBMM and the PIC-MCC solver is obtained for $t \leq 5$\unit{\micro\second} for the temperature and $t \leq 1$\unit{\micro\second} for the excess kurtosis, and for all times for $M_2$ and $M_4$. 
    However, electrons are thermalized over time in the QBMM case due to the use of the BGK operator for elastic electron-neutral collisions.
    As a result, $\Delta \rightarrow 0$ as $t$ increases, the high-energy tail of the VDF is not captured correctly for $t\geq 5$\unit{\micro\second}, and the 
    final temperature reported by QBMM is lower than that of the PIC-MCC approach.
    This, in turn, reduces the calculations of the collision reaction rate. 
    Additional collision processes would further enhance this effect. 

    For this particular example, additional HyQMOM nodes do not improve the results as the limiting numerical phenomenon
    is associated with the BGK operator.
    A possible remedy consists of removing the assumption of transverse equilibrium through   
    the implementation of a QBMM that is able to represent \textit{all} three velocity directions 
    (\textit{e.g.}, 3-D Conditional HyQMOM (CHyQMOM) with 16 moments\cite{Patel2019} or 3-D HyQMOM with 35 moments\cite{Bryngelson2025}), 
    and to consider a more general collision operator for elastic collisions.
    This is beyond the scope of this manuscript and is reserved for future work.

    \clearpage
\newpage
	\subsection{RF capacitive discharge}
    
	We finally turn our attention to the helium RF discharge test cases 
    described in details by Turner~\textit{et al.}~\cite{Turner2013}
	The benchmark represents a one-dimensional capacitive RF discharge between two electrodes and 
    includes collisions between electrons, ions,
	and neutrals. 
    The collisions considered are electron-neutral, both elastic and inelastic (impact-ionization, and two excitation collisions),
    as well as elastic ion-neutral. The elastic ion-neutral collisions are split between an isotropic and a backscattering
    component.
    The electron-neutral and ion-neutral cross sections are sourced from the Biagi v7.1\cite{LXCAT-Biagi} and Phelps\cite{LXCAT-Phelps} databases, respectively.
    Both electrons and ions are evolved in time in a set background of neutrals.
    
    We consider only the ``test case 1'', which has the lowest neutral pressure ($P_n = 1$~mTorr). The simulation parameters
    are shown in Table~\ref{tbl:turner-benchmark}.
    This is a challenging case for QBMM: 
    there are no Coulomb collisions and fewer collisions with neutrals as compared to higher-pressure cases 
    that may bring the electron and ion VDF back towards equilibrium. 
    The presence of boundaries also results in a mostly collisionless sheath wherein the moment set is 
    brought towards the edge of the realizability domain. 
    In particular, numerical accuracy may bring an otherwise realizable moment set over the realizability boundary. 
    In those instances, higher-order moments ($M_3$ and $M_4$) are adjusted such that the moment set remains realizable. 

    \begin{table}[ht!]
        \centering
\begin{ruledtabular}
        \begin{tabular}{lll}
        Quantity & Species & Value \\
        \hline
        Initial temperatures (K)                        & Electrons     & 30,000 \\
                                                        & Ions, neutrals          & 300\\
        Initial density (m\textsuperscript{-3})         & Electrons, ions       & 2.56$\times 10^{14}$ \\
                                                        & Neutrals              & 9.67$\times 10^{20}$\\
        \hline
        RF frequency (MHz), $f_{RF}$  & --- & 13.56 \\
        RF voltage (V)      & --- & 450 \\
        \hline
        Domain length (cm)  & --- & 6.7 \\
        Simulation time (\SI{ }{\micro\second}) & --- & 3 \\
        \end{tabular}
\end{ruledtabular}
        \caption{Parameters for the capacitive discharge benchmark.}
        \label{tbl:turner-benchmark}
    \end{table}
    
    As discussed in the previous test case, the BGK operator chosen for electron-neutral elastic collisions 
    artificially restores the electron VDF towards Maxwellian. As a result, the ionization reaction rate is either under or over-estimated.
    For this particular case, much like the previous example, the electron temperature is underestimated, which
    leads to a decrease in the electron density for longer simulation times such that the overall discharge extinguishes itself. 
    It is informative, however, to compare instantaneous snapshots of the QBMM and PIC-MCC solutions to demonstrate
    the capabilities of QBMM for moderate deviations from equilibrium. We do so
    for up to \SI{3}{\micro\second} (\textit{i.e.}, $\approx 40$~RF cycles).

    \begin{table}[ht!]
    \centering
    \begin{ruledtabular}
        \begin{tabular}{lll}
        Algorithm     &  Quantity & Value\\
        \hline
        PIC           & Particles per cell & 2048\\
                      & Number of cells & 128 \\
                      & Timestep        & 1/400$\cdot f_{RF}$\\
       \hline
        QBMM          & Number of cells & 2048 \\
                      & Flux type & Kinetic \\
                      & EQMOM nodes for collision integrals & 2 \\
                      & HyQMOM nodes for transport (e\textsuperscript{-}, He\textsuperscript{+}) & 3\\
                      & CFL number & 0.5
        \end{tabular}
    \end{ruledtabular}
    \caption{Domain discretization and algorithm parameters for the PIC and QBMM approaches.}
    \label{tbl:turner-discretization}
    \end{table}

    The salient algorithm parameters and domain discretization are shown in Table~\ref{tbl:turner-discretization}.
    As compared to Turner~\textit{et al.}, we increase the total number of particles per cell to 2,048 to 
    reduce statistical noise.
    The QBMM solution is computed on a spatial grid with a 2,048 cells.
    Because the QBMM flux operator is  
    of first order and, therefore, highly diffusive, 
    a large number of cells is required in order
    to avoid diffusion in physical space. 
    For example, we have observed that the second-order moment is over-estimated   
    for a grid with 128 cells, which, in turn, yields a higher-than-normal electron temperature and, therefore,
    an increase in ionization rate. This increased ionization rate then further increases the second-order moment in
    a positive feedback loop.
    
    We limit our study to three nodes only. The solution for a higher number of nodes is highly sensitive to the evaluation of 
    reaction integrals of higher order (Equation~\ref{eqn:reaction-rate-inelastic} with $k \geq 3$ with the integration
    kernels shown in the Appendix), especially for ionization, which has a high threshold energy.

    In PIC-MCC, the boundary condition for the wall is such that particles that reach the boundary are lost without replacement.
    Because we use a finite volume approach for QBMM, the flux must be evaluated at the edge of the cell that is adjacent
    to the physical boundary. 
    We simply set the VDF to zero at the boundary (\textit{e.g.}, $\tilde{f}_{i+1/2} = 0$ for the right-hand side boundary located next to cell $i$). 
    For example, for the right-hand side boundary, the kinetic flux, $F^k_{i+1/2}$, 
    for the $k$-th moment at the edge of cell $i$ is given by (with flux splitting):
    \begin{equation}
    \begin{aligned}
        F^k_{i+1/2} &= 
       \int_0^{+\infty} \tilde{f}_i\left(x,v\right)v^{k+1} \diff v 
       + 
       \int_{-\infty}^0 \tilde{f}_{i+1}\left(x,v\right) v^{k+1}\diff v \\ 
       &=  
       \int_0^{+\infty} \tilde{f}_i\left(x,v\right)v^{k+1} \diff v. 
    \end{aligned}
    \end{equation}

\begin{figure*}
		\centering
		\includegraphics[scale=0.9]{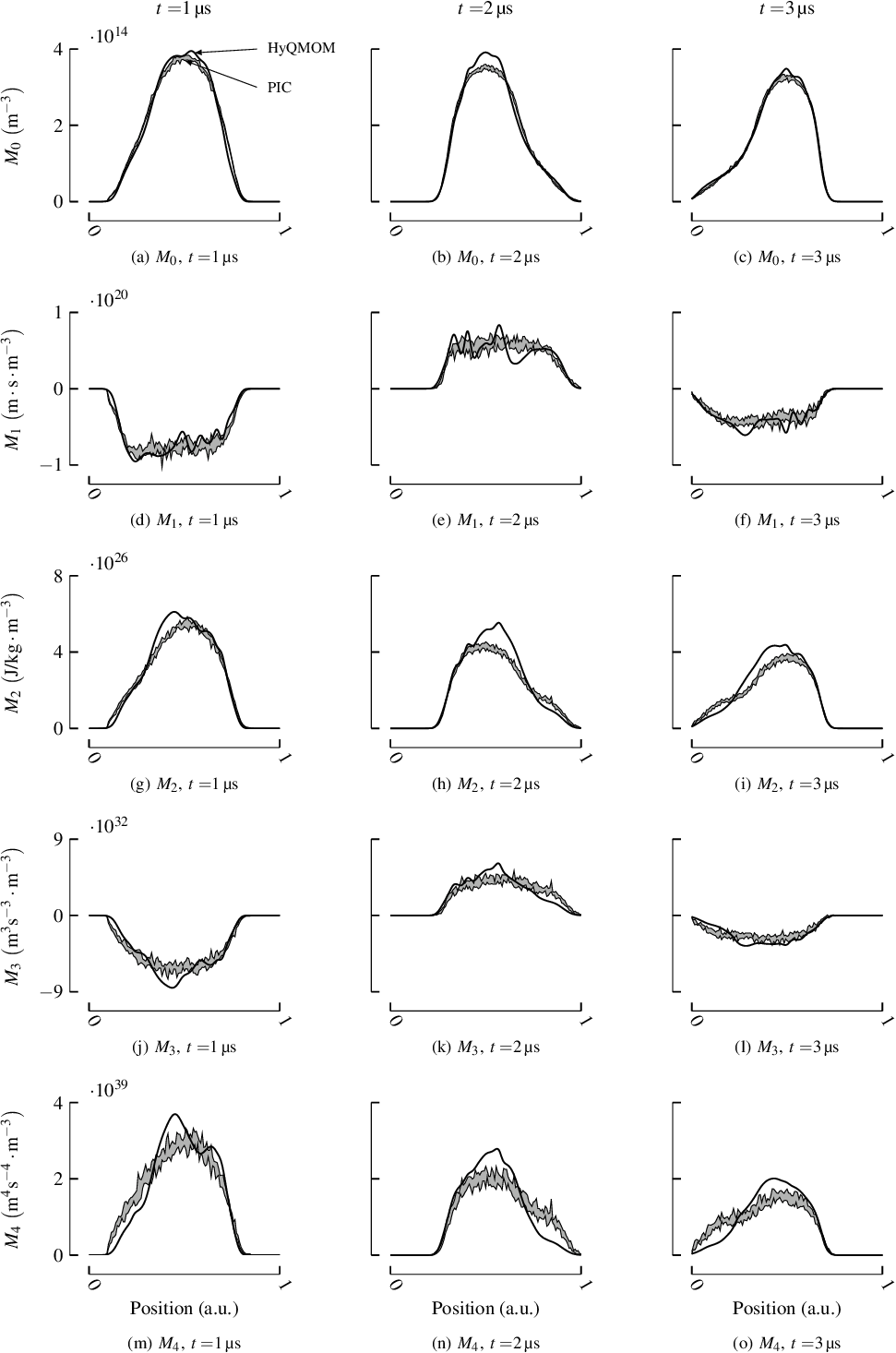}
		\caption{Moments of the electron VDF in the $x$-direction as computed with PIC and QBMM for $t=$1--3~\SI{}{\micro\second}. 
        (a--c) Density ($M_0$), (d--f) momentum ($M_1$), (g--i) energy ($M_2$), (j--l) skewness ($M_3$), and (m--o) kurtosis ($M_4$). 
        }
		\label{fig:turner-et-al-results-moments}
	\end{figure*}

\begin{figure*}
		\centering
		\includegraphics[scale=0.9]{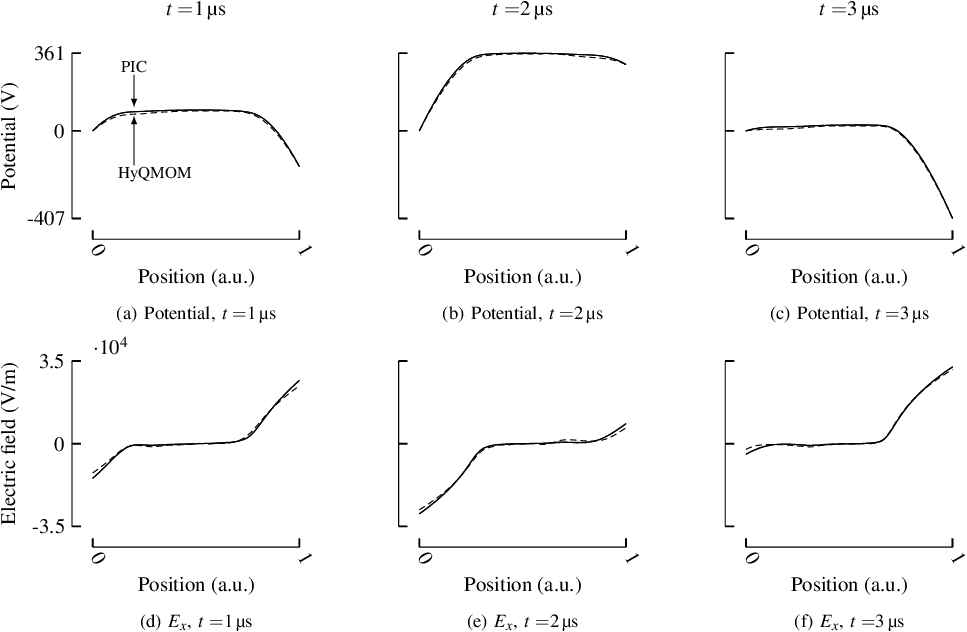}
		\caption{Electric (a--c) potential and (d--f) field as computed with PIC and QBMM for $t=$1--3~\SI{}{\micro\second}.} 
		\label{fig:turner-et-al-results-electrical}
    \end{figure*}

	Figure~\ref{fig:turner-et-al-results-moments} 
    shows
    the moments of
    the electron species
    as  
    obtained through PIC and QBMM,
    at three different times, $t\in \left\{1,2,3\right\}$~\unit{\micro\second}. 
    Potential and electric field are shown in Figure~\ref{fig:turner-et-al-results-electrical}.
    Good qualitative and quantitative agreement is obtained for all quantities. The oscillations that appear in the QBMM solution 
    have been observed in other moderately collisional problems (see, \textit{e.g.}, Ref.~\onlinecite{Taunay2023}).
    Past \SI{3}{\micro\second}, the artificial restoration towards equilibrium
    due to the elastic electron-neutral BGK operator affects the underlying VDF, and, therefore, the integrated moments.

   \begin{figure*}
		\centering
		\includegraphics[scale=0.7]{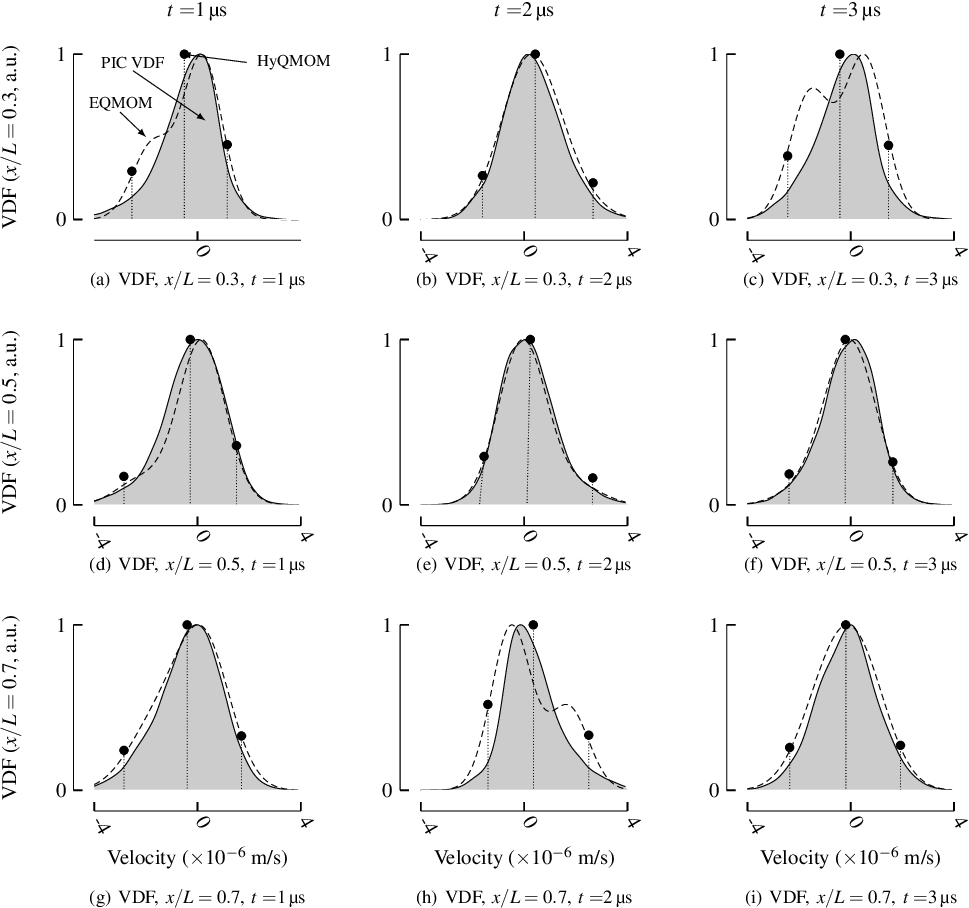}
		\caption{Electron VDF in the parallel direction ($x$) 
        as obtained from PIC and QBMM, at three different locations. (a--c) $x/L = 0.3$, (d--f) $x/L=0.5$, and (g--i) $x/L=0.7$.
        Both approximate VDFs used for transport (three-node HyQMOM) and for reaction rate calculation (two-node EQMOM) are displayed.
        The VDF is normalized by its largest value.
        }
		\label{fig:turner-et-al-results-VDF-parallel}
	\end{figure*}

    \begin{figure*}
		\centering
		\includegraphics[scale=0.8]{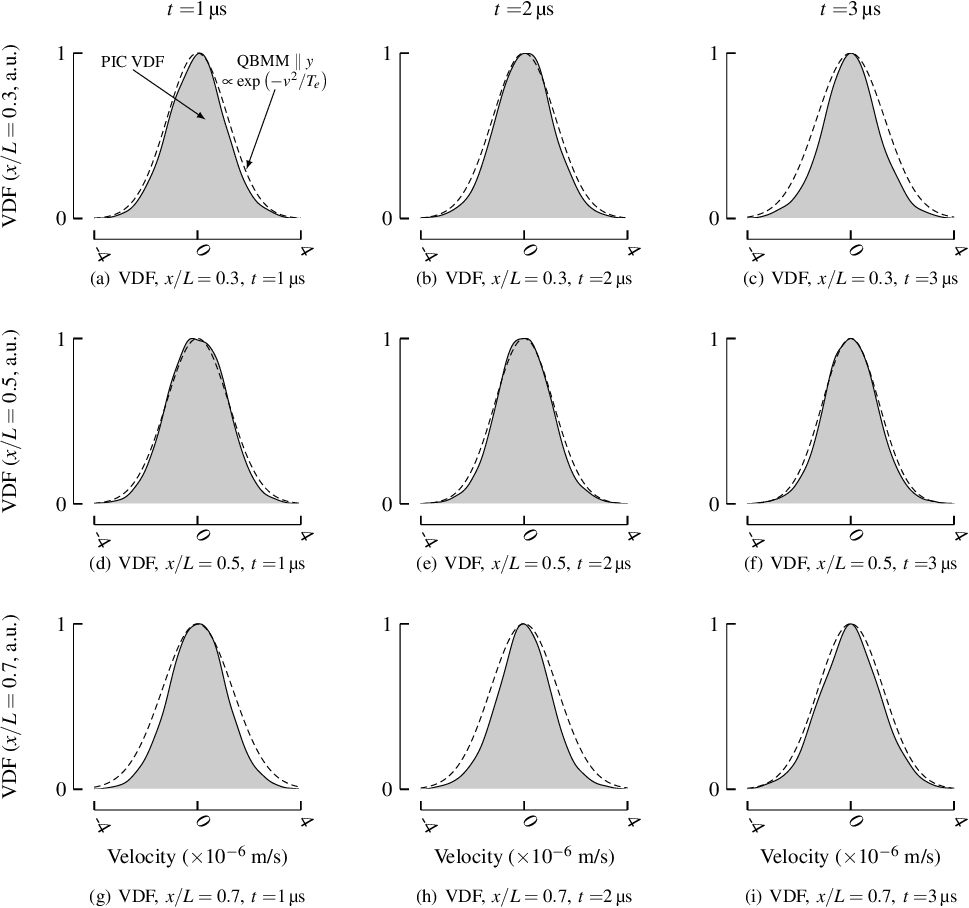}
        \caption{Electron VDF in the transverse direction ($yz$) 
        as obtained from PIC and QBMM, at three different locations. (a--c) $x/L = 0.3$, (d--f) $x/L=0.5$, and (g--i) $x/L=0.7$.
        The VDF is normalized by its largest value.
        }
		\label{fig:turner-et-al-results-VDF-transverse}
	\end{figure*}
    
    The VDF in the parallel ($x$) and transverse ($yz$) directions is shown at three positions ($x\in \left\{0.3,0.5,0.7\right\}$) 
    and at three different times ($t\in \left\{1,2,3\right\}$~\unit{\micro\second}) in Figures~\ref{fig:turner-et-al-results-VDF-parallel}
    and~\ref{fig:turner-et-al-results-VDF-transverse}, respectively. 
    Both the HyQMOM representation of the parallel-direction VDF that is used for transport and the corresponding EQMOM representation 
    that is used to compute the reaction rates are shown in Figure~\ref{fig:turner-et-al-results-VDF-parallel}.
    The VDF is adequately estimated at the center of the domain ($x/L = 0.5$). However,
    near the sheath entrance ($x/L=0.3$ and $0.7$), HyQMOM fails to capture the correct location of the peak
    of the VDF, and EQMOM produces a VDF that is too broad and that may feature a secondary peak. 
    This, in turn, leads to an overestimation of the reaction rates, and, therefore, an inaccurate estimation of the
    moments.
    Additional nodes (\textit{e.g.}, five nodes) would be beneficial. For this problem, however, additional nodes result in
    an unstable solution, which may be due to  
    the underlying flux calculations
    that results in ill-conditioned Vandermonde matrices to find the weights and abscissas, 
    or numerical accuracy of the the higher-order reaction ionization integral for non-Maxwellian, drifting kernels. 
    
    The VDF in the transverse direction (Figure~\ref{fig:turner-et-al-results-VDF-transverse}) remains near-Maxwellian and is well-resolved in most cases, even with the simplifying
    \textit{ansatz} of thermal equilibrium in 
    Equation~\ref{eqn:simplifying-transverse-equilibrium}.
    Provided that the electron temperature as obtained from QBMM is close to that predicted by PIC, the transverse VDF is
    adequately captured.

\section{Conclusion}
	We have developed the mathematical framework for collisional processes relevant to low-temperature plasmas in QBMM. 
	We computed the moments of the collision terms for both the BGK and Boltzmann operators for elastic and inelastic
	threshold collisions, respectively, validated the implementation against several benchmark cases, and explored
    the limitations of QBMM. 
    We showed both qualitative and quantitative 
	agreement with both particle methods and other moment methods, which indicates that QBMM are, in most cases, able to accurately represent the
	underlying physics of the simulated problems.
	
    Several avenues for improvement have been identified. The application of QBMM was limited to a single velocity direction
    and equilibrium was assumed in the transverse direction, as seen in the \textit{ansatz} for the VDF (Equation~\ref{eqn:simplifying-transverse-equilibrium}).
    However, this work may be generalized to multiple velocity directions by using multi-dimensional QBMM for both EQMOM and
    HyQMOM.\cite{Patel2019,Bryngelson2025}
    EQMOM is preferred to compute the reaction integrals as those can be simplified from 6D to 1D when using 
    exponential kernels.
    Boundary conditions beyond simple loss of particles can be implemented in this framework, as long as the VDF is known at the edges of the domain. 
    For example, in 2D2V, a specular reflection boundary condition in CHyQMOM has been implemented by the author in separate work.\cite{Taunay2022} 
	The finite-volume kinetic flux\cite{Fox2018} used to transport 
	the moments in 1-D was kept simple in this work.
	The HLL flux as suggested by Fox and Laurent\cite{Fox2021} 
	might improve the accuracy of the solution as it is less diffusive than the kinetic flux.
    These limitations will be addressed in future work.

	We have demonstrated that QBMM are able to reproduce benchmark results from the
low-temperature plasma literature. The addition of multi-species collisions enhances the original
collisionless approach from Taunay and Mueller\cite{Taunay2023} and enables the expedient and noise-free simulation
of low-temperature plasmas relevant to plasma propulsion and semiconductor processing. 
	The mathematical framework developed in this work may be generalized to higher dimensions and also to additional collision operators that
are relevant to low-temperature plasmas. 
	A complete set of collisions coupled to time-dependent boundary conditions
	may enable QBMM to efficiently simulate whole devices that feature complex chemistry and that span 
    multiple flow regimes (\textit{e.g.}, from continuum to rarefied flow),
    at engineering space- and time-scales.

%%%%%%%%%%%%%%%%%%%%%%%%%%%%%%%%%%%%%%%%%%%%%%%%%%%%%%%%%%%%%%%%%%%%%%%%%%%%%%%%%%%%%%%%%%%%%%%%%%%%%%%%%%%%%%%%%

\section*{Acknowledgments}
This work was supported AFOSR (Propulsion and Power program).
The views expressed in this article are those of the author and do not reflect the official policy or position of the U.S. Naval Academy, Department of the Navy, Department of Defense, the U.S. Government, or any agency thereof.

The author thanks Christopher Wordingham and Willca Villafana for fruitful discussions.

\section*{Author declaration}
\subsection*{Data availability}
The data that support the findings of this study are available from the corresponding author
upon reasonable request.

\subsection*{Conflict of interest}
The author has no conflicts to disclose.

\section*{Appendix}
\setcounter{equation}{0}

\subsection*{A. Integrals of drifting bi-Maxwellian distributions}

The 6D collision integral over velocity space that arises in the computation of the reaction rates may be simplified into a single
1D integral over energy space when the two colliding particles follow a Maxwellian distribution that may be anisotropic (\textit{i.e.}, $T_\perp \neq T_\parallel$) and/or drifting.
The 1D integrals are presented by Xie~\textit{et al.}\cite{Xie2023} for four different cases in which the colliding particles 
$s=\{i,j\}$ 
follow 
a Maxwellian VDF that is either
\begin{enumerate}
    \item isothermal:
    \begin{equation}
  f_s\left(\textrm{v}\right) = \left(\dfrac{m_s}{2\pi k_B T_s}\right)^{3/2} \exp\left(-\dfrac{m_s v^2}{2 k_B T_s}\right)  
\end{equation}
    \item isothermal and drifting at velocity $\mathbf{v_{d,s}}$: 
\begin{equation}
  f_s\left(\textrm{v}\right) = \left(\dfrac{m_s}{2\pi k_B T_s}\right)^{3/2} \exp\left(-\dfrac{m_s \left(\mathbf{v}-\mathbf{v_{d,s}}\right)^2}{2 k_B T_s}\right), 
\end{equation}
    \item anisotropic with parallel (resp. perpendicular) velocity and temperature $v_{\parallel}$ and $T_{\parallel,s}$ (resp. $v_{\perp}$ and $T_{\perp,s}$): 
\begin{align}
  f_s\left(\textrm{v}\right) &= 
  \dfrac{1}{T_{\parallel,s}^{1/2}T_{\perp,s}}
  \left(\dfrac{m_s}{2\pi k_B}\right)^{3/2}\times\dots\\ 
  &\dots \exp\left(-\dfrac{m_s v_\perp^2}{2 k_B T_{\perp,s}}-\dfrac{m_s v_\parallel^2}{2 k_B T_{\parallel,s}}\right), 
\end{align}
    \item anisotropic with a drifting velocity $v_{d,s}$ along the parallel direction: 
    \begin{align}
f_s\left(\textrm{v}\right) &= 
\dfrac{1}{T_{\parallel,s}^{1/2}T_{\perp,s}}
\left(\dfrac{m_s}{2\pi k_B}\right)^{3/2}\times\dots\\ 
&\dots\exp\left(-\dfrac{m_s v_\perp^2}{2 k_B T_{\perp,s}}-\dfrac{m_s \left(v_\parallel-v_{d,s}\right)^2}{2 k_B T_{\parallel,s}}\right), 
    \end{align}
\end{enumerate}
For the anisotropic cases, the total temperature is
\begin{equation}
    T_s = \dfrac{1}{3}\left(2 T_{\perp,s} + T_{\parallel,s}\right).
\end{equation}
In all cases, the reaction rate for the collision of species $i$ and $j$ is given by
\begin{equation}
    R_{ij} = \int_{0}^{+\infty} Q\left(E\right) K\left(R_t, R_d\right) \diff{E}, 
\end{equation}
where $Q$ is the collision cross section
and $K$ is a kernel function that depends on the shape of the VDF.
The parameters $R_t$ and $R_d$ are defined as
\begin{gather}
    R_t = \dfrac{T_{\perp,r}}{T_{\parallel,r}},\,\textrm{and}\\
    R_d = \dfrac{E_d}{k_B T_r},
\end{gather}
where $E_d = k_B T_d = \dfrac{m_r v_d^2}{2}$, $v_d = \left|\mathbf{v_{d,i}}-\mathbf{v_{d,j}}\right|$.
The reduced temperature and mass, $T_r$ and $m_r$, are defined as
\begin{equation}
    T_r = \dfrac{m_i T_j + m_j T_i}{m_i + m_j} 
\end{equation}
and 
\begin{equation}
    m_r = \dfrac{m_i m_j}{m_i + m_j}
\end{equation}
respectively.
Table~\ref{tbl:kernels} summarizes the different kernels.

\begin{table*}[t]
\caption{\label{tbl:kernels}Kernel formulation for the calculation of the reaction rate of two species $i$ and $j$ that follow
a Maxwellian distribution that is possibly anisotropic and/or drifting.}
\begin{ruledtabular}
\begin{tabular}{ll}
Case & Kernel\\
Isothermal & $\sqrt{\dfrac{8}{\pi m_r}}\dfrac{1}{\left(k_B T_r\right)^{3/2}} E \exp\left(-\dfrac{E}{k_B T_r}\right)$\\
Isothermal, drifting & 
   $\sqrt{\dfrac{2}{\pi m_r k_B^2 T_r^2}}
    \dfrac{\exp\left(-R_d\right)}{\sqrt{R_d}}
    \sqrt{E} 
    \exp\left(-\dfrac{E+E_d}{k_B T_r}\right) 
    \sinh\left(\dfrac{2\sqrt{E}}{k_B T_r}\sqrt{R_d}\right)
    $ \\
Anisotropic & 
$\sqrt{\dfrac{1}{9 m_r k_B^2 T_r^2}}
\sqrt{\dfrac{\left(2 R_t+1\right)^2}{R_t\left(R_t-1\right)}}
\sqrt{E}
\exp\left(-\dfrac{E}{k_B T_r}\dfrac{2 R_t+1}{3 R_t}\right) 
\textrm{erf}\left(\sqrt{\dfrac{E}{k_B T_r}\dfrac{\left(R_t-1\right)\left(2 R_t+1\right)}{3 R_t}}\right)
$
\\
Anisotropic, drifting & 
$\sqrt{\dfrac{1}{18 m_r k_B^2 T_r^2}}
\exp\left(\dfrac{R_d}{R_t-1}\dfrac{2 R_t+1}{3 R_t}\right)
\sqrt{\dfrac{\left(2 R_t+1\right)^2}{R_t\left(R_t-1\right)}}
\sqrt{E}
\exp\left(-\dfrac{E}{k_B T_r}\dfrac{2 R_t+1}{3 R_t}\right)\times \dots 
$\\
& 
$\dots\left[
\textrm{erf}
\left(
\sqrt{\dfrac{E}{k_B T_r} \dfrac{\left(R_t-1\right)\left(2 R_t+1\right)}{3 R_t} }
+
\sqrt{\dfrac{R_d R_t}{R_t-1}\dfrac{2 R_t+1}{3}}
\right)
+
\textrm{erf}
\left(
\sqrt{\dfrac{E}{k_B T_r} \dfrac{\left(R_t-1\right)\left(2 R_t+1\right)}{3 R_t} }
-
\sqrt{\dfrac{R_d R_t}{R_t-1}\dfrac{2 R_t+1}{3}}
\right)
\right]
$
\\
\end{tabular}
\end{ruledtabular}
\end{table*}

The numerical evaluation of the reaction rate for the anisotropic, drifting kernel (case 4) necessitates additional care:
an indeterminate numerical number (NaN) is typically obtained for $E/k_B T_r < \dfrac{R_d}{R_t-1}$ as 
the two error functions cancel out while the exponential factor quickly grows.
In practice, if $\left|R_t-1\right|\leq 0.1$, we assume that $T_\perp\approx T_\parallel$ and use the isotropic, drifting kernel. 
Large values of $R_d$ have been observed with a vanishingly small weight in the EQMOM (\textit{i.e.}, $\omega_\alpha \rightarrow 0$ while $v_\alpha \rightarrow +\infty$). To avoid divergent numerical integrals, the reaction rate is clamped to a zero value for $R_d > 20$ (\textit{i.e.}, twenty times the thermal energy). 
Because the kernels depend on a finite set of parameters that can be easily bounded ($T_r$, $E_d$, $R_t$, $R_d$), all
collision integrals can be performed separately and stored as a look-up table, thus improving
computation time.

\section*{References}

\bibliography{biblio}

@book{Zhdanov,
	title={Transport processes in multicomponent plasma},
	author={Zhdanov, V. M.},
	year={2002},
	publisher={Taylor \& Francis}
}

@book{Vincenti,
title={Introduction to physical gas dynamics},
author={Vincenti, Walter G. AND Kruger, Charles Jr. H.},
year={1965},
publisher={John Wiley \& Sons}
}

@book{NRLFormulary,
	title={{NRL} plasma formulary},
	author={Huba, Joseph Donald},
	volume={6790},
	number={98-358},
	year={1998},
	publisher={Naval Research Laboratory}
}

@misc{A276985,
	author={{OEIS Foundation Inc.}},
	year={2025},
	title={Entry {A276985} in the On-Line Encyclopedia of Integer Sequences},
	note={\url{https://oeis.org/A276985}}
}

@misc{NIST:DLMF,
    title = {{NIST} Digital Library of Mathematical Functions --- Table 3.5.17.5},
    howpublished = {\url{https://dlmf.nist.gov/3.5#T17_5}, Release 1.2.7 of 2026-06-15},
    note = {{F.~W.~J. Olver, A.~B. {Olde Daalhuis}, D.~W. Lozier, B.~I. Schneider,
                R.~F. Boisvert, C.~W. Clark, B.~R. Miller, B.~V. Saunders,
                H.~S. Cohl, and M.~A. McClain, eds.}}
                }

@article{SciPy2020,
  author  = {Virtanen, Pauli and Gommers, Ralf and Oliphant, Travis E. and
            Haberland, Matt and Reddy, Tyler and Cournapeau, David and
            Burovski, Evgeni and Peterson, Pearu and Weckesser, Warren and
            Bright, Jonathan and {van der Walt}, St{\'e}fan J. and
            Brett, Matthew and Wilson, Joshua and Millman, K. Jarrod and
            Mayorov, Nikolay and Nelson, Andrew R. J. and Jones, Eric and
            Kern, Robert and Larson, Eric and Carey, C J and
            Polat, {\.I}lhan and Feng, Yu and Moore, Eric W. and
            {VanderPlas}, Jake and Laxalde, Denis and Perktold, Josef and
            Cimrman, Robert and Henriksen, Ian and Quintero, E. A. and
            Harris, Charles R. and Archibald, Anne M. and
            Ribeiro, Ant{\^o}nio H. and Pedregosa, Fabian and
            {van Mulbregt}, Paul and {SciPy 1.0 Contributors}},
  title   = {{{SciPy} 1.0: Fundamental Algorithms for Scientific
            Computing in Python}},
  journal = {Nature Methods},
  year    = {2020},
  volume  = {17},
  pages   = {261--272},
  adsurl  = {https://rdcu.be/b08Wh},
  doi     = {10.1038/s41592-019-0686-2},
}

@book{Piessens2012,
  title={Quadpack: a subroutine package for automatic integration},
  author={Piessens, Robert and de Doncker-Kapenga, Elise and {\"U}berhuber, Christoph W and Kahaner, David K},
  year={2012},
  publisher={Springer Science \& Business Media}
}

@article{Janes1966,
  title={Anomalous electron diffusion and ion acceleration in a low-density plasma},
  author={Janes, GS and Lowder, RS},
  journal={The Physics of Fluids},
  volume={9},
  number={6},
  pages={1115--1123},
  year={1966},
  publisher={American Institute of Physics}
}

@article{Meezan2001,
  title={Anomalous electron mobility in a coaxial Hall discharge plasma},
  author={Meezan, Nathan B and Hargus Jr, William A and Cappelli, Mark A},
  journal={Physical Review E},
  volume={63},
  number={2},
  pages={026410},
  year={2001},
  publisher={APS}
}

@article{Mikellides2007,
  title={Evidence of nonclassical plasma transport in hollow cathodes for electric propulsion},
  author={Mikellides, Ioannis G and Katz, Ira and Goebel, Dan M and Jameson, Kristina K},
  journal={Journal of Applied Physics},
  volume={101},
  number={6},
  year={2007},
  publisher={AIP Publishing}
}

@article{Mikellides2005,
  title={{Hollow cathode theory and experiment. II. A two-dimensional theoretical model of the emitter region}},
  author={Mikellides, Ioannis G and Katz, Ira and Goebel, Dan M and Polk, James E},
  journal={Journal of Applied Physics},
  volume={98},
  number={11},
  year={2005},
  publisher={AIP Publishing}
}

@article{Ortega2018,
  title={Hollow cathode simulations with a first-principles model of ion-acoustic anomalous resistivity},
  author={Ortega, Alejandro Lopez and Jorns, Benjamin A and Mikellides, Ioannis G},
  journal={Journal of Propulsion and Power},
  volume={34},
  number={4},
  pages={1026--1038},
  year={2018},
  publisher={American Institute of Aeronautics and Astronautics}
}

@article{Sary2017,
  title={Hollow cathode modeling: I. A coupled plasma thermal two-dimensional model},
  author={Sary, Ga{\'e}tan and Garrigues, Laurent and Boeuf, Jean-Pierre},
  journal={Plasma Sources Science and Technology},
  volume={26},
  number={5},
  pages={055007},
  year={2017},
  publisher={IOP Publishing}
}

@article{Broadwell1964,
  title={Study of rarefied shear flow by the discrete velocity method},
  author={Broadwell, James E},
  journal={Journal of Fluid Mechanics},
  volume={19},
  number={3},
  pages={401--414},
  year={1964},
  publisher={Cambridge University Press}
}

@article{Xu2010,
  title={A unified gas-kinetic scheme for continuum and rarefied flows},
  author={Xu, Kun and Huang, Juan-Chen},
  journal={Journal of Computational Physics},
  volume={229},
  number={20},
  pages={7747--7764},
  year={2010},
  publisher={Elsevier}
}

@article{Juno2018,
  title={Discontinuous Galerkin algorithms for fully kinetic plasmas},
  author={Juno, James and Hakim, Ammar and TenBarge, Jason and Shi, Eric and Dorland, William},
  journal={Journal of Computational Physics},
  volume={353},
  pages={110--147},
  year={2018},
  publisher={Elsevier}
}

@article{Raisanen2019,
  title={Two-dimensional hybrid-direct kinetic simulation of a Hall thruster discharge plasma},
  author={Raisanen, Astrid L and Hara, Kentaro and Boyd, Iain D},
  journal={Physics of Plasmas},
  volume={26},
  number={12},
  year={2019},
  publisher={AIP Publishing}
}

@book{Bird1994,
author={Bird, G. A.},
title={Molecular gas dynamics and the direct simulation of gas flows},
year={1994},
}

@book{Birdsall2004,
author={Birdsall, C. K. AND Langdon, A. B.},
title={Plasma physics via computer simulation},
year={2004},
publisher={CRC Press}}

@article{Nanbu1997,
  title={Theory of cumulative small-angle collisions in plasmas},
  author={Nanbu, K},
  journal={Physical Review E},
  volume={55},
  number={4},
  pages={4642},
  year={1997},
  publisher={APS}
}

@article{Sahu2020,
  title={Full fluid moment model for low temperature magnetized plasmas},
  author={Sahu, Rupali and Mansour, Adnan R and Hara, Kentaro},
  journal={Physics of Plasmas},
  volume={27},
  number={11},
  year={2020},
  publisher={AIP Publishing}
}

@article{Patel2019,
  title={Three-dimensional conditional hyperbolic quadrature method of moments},
  author={Patel, Ravi G and Desjardins, Olivier and Fox, Rodney O},
  journal={Journal of Computational Physics: X},
  volume={1},
  pages={100006},
  year={2019},
  publisher={Elsevier}
}

@article{Bryngelson2025,
  title={Fourth-Order HyQMOM Closures for Multidimensional Kinetic Equations},
  author={Bryngelson, Spencer and Fox, Rodney O and Laurent, Fr{\'e}d{\'e}rique},
  note={Preprint. \url{https://hal.science/hal-05398171v1}},
  year={2025}
}

@article{Fox2023,
	title={The generalized quadrature method of moments},
	author={Fox, Rodney O and Laurent, Fr{\'e}d{\'e}rique and Passalacqua, Alberto},
	journal={Journal of Aerosol Science},
	volume={167},
	pages={106096},
	year={2023},
	publisher={Elsevier}
}

@article{Turner2013,
	title={Simulation benchmarks for low-pressure plasmas: Capacitive discharges},
	author={Turner, Miles M and Derzsi, Aranka and Donko, Zoltan and Eremin, Denis and Kelly, Sean J and Lafleur, Trevor and Mussenbrock, Thomas},
	journal={Physics of Plasmas},
	volume={20},
	number={1},
	year={2013},
	publisher={AIP Publishing}
}

@article{Bhatnagar1954,
	title={{A model for collision processes in gases. I. Small amplitude processes in charged and neutral one-component systems}},
	author={Bhatnagar, Prabhu Lal and Gross, Eugene P and Krook, Max},
	journal={Physical Review},
	volume={94},
	number={3},
	pages={511},
	year={1954},
	publisher={APS}
}

@article{Haack2017,
	title={A conservative, entropic multispecies BGK model},
	author={Haack, Jeffrey R and Hauck, Cory D and Murillo, Michael S},
	journal={Journal of Statistical Physics},
	volume={168},
	number={4},
	pages={826--856},
	year={2017},
	publisher={Springer}
}

@article{Alvarez2022,
	title={A regularized high-order moment model to capture non-Maxwellian electron energy distribution function effects in partially ionized plasmas},
	author={Alvarez Laguna, A and Esteves, B and Bourdon, A and Chabert, P},
	journal={Physics of Plasmas},
	volume={29},
	number={8},
	year={2022},
	publisher={AIP Publishing}
}

@article{Alvarez2026,
  title={High-order moment closure for nonmagnetized electrons in partially ionized plasmas},
  author={Alvarez Laguna, A and Hara, K},
  journal={Physical Review E},
  volume={113},
  number={2},
  pages={025207},
  year={2026},
  publisher={APS}
}

@article{Wu2026,
title={Analysis of non-local heat flux in capacitively coupled plasmas},
author={Wu, Geyi AND Yamashita, Y. AND Mansour, A.R. AND Alvarez Laguna, A. AND Hara, K},
journal={Plasma Sources Science and Technology },
volume={35},
number={2},
pages={025032},
year={2026}
}

@article{Taunay2023,
	title={Quadrature-based moment methods for kinetic plasma simulations},
	author={Taunay, Pierre-Yves C R and Mueller, Michael E},
	journal={Journal of Computational Physics},
	volume={473},
	pages={111700},
	year={2023},
	publisher={Elsevier}
}

@article{Taunay2022,
	title={Quadrature-based moment methods applied to E$\times$B discharges},
	author={Taunay, Pierre-Yves C R and Mueller, Michael E},
	journal={37th International Electric Propulsion Conference},
	year={2022},
	note={IEPC-2022-309}
}

@article{Janhunen2018,
	title={Nonlinear structures and anomalous transport in partially magnetized E$\times$ B plasmas},
	author={Janhunen, Salomon and Smolyakov, Andrei and Chapurin, Oleksandr and Sydorenko, Dmytro and Kaganovich, Igor and Raitses, Yevgeni},
	journal={Physics of Plasmas},
	volume={25},
	number={1},
	pages={011608},
	year={2018},
	publisher={AIP Publishing LLC}
}

@article{Hara2012,
	title={One-dimensional hybrid-direct kinetic simulation of the discharge plasma in a Hall thruster},
	author={Hara, Kentaro and Boyd, Iain D and Kolobov, Vladimir I},
	journal={Physics of Plasmas},
	volume={19},
	number={11},
	pages={113508},
	year={2012},
	publisher={American Institute of Physics}
}

@article{Langdon1979,
	title={Kinetic theory for fluctuations and noise in computer simulation of plasma},
	author={Langdon, A Bruce},
	journal={The Physics of Fluids},
	volume={22},
	number={1},
	pages={163--171},
	year={1979},
	publisher={American Institute of Physics}
}

@article{Grad1949,
	title={On the kinetic theory of rarefied gases},
	author={Grad, Harold},
	journal={Communications on Pure and Applied Mathematics},
	volume={2},
	number={4},
	pages={331--407},
	year={1949},
	publisher={Wiley Online Library}
}

@article{Struchtrup2003,
  title={Regularization of Grad’s 13 moment equations: Derivation and linear analysis},
  author={Struchtrup, Henning and Torrilhon, Manuel},
  journal={Physics of Fluids},
  volume={15},
  number={9},
  pages={2668--2680},
  year={2003},
  publisher={American Institute of Physics}
}

@article{Torrilhon2004,
  title={Regularized 13-moment equations: shock structure calculations and comparison to Burnett models},
  author={Torrilhon, Manuel and Struchtrup, Henning},
  journal={Journal of Fluid Mechanics},
  volume={513},
  pages={171--198},
  year={2004},
  publisher={Cambridge University Press}
}

@article{kuldinow2024-2,
  title={Ten-moment fluid model for low-temperature magnetized plasmas},
  author={Kuldinow, Derek Amur and Yamashita, Yusuke and Hara, Kentaro},
  journal={Physics of Plasmas},
  volume={31},
  number={12},
  year={2024},
  publisher={AIP Publishing}
}

@article{Kuldinow2024-1,
  title={Ten-moment fluid model with heat flux closure for gasdynamic flows},
  author={Kuldinow, Derek A and Yamashita, Yusuke and Mansour, Adnan R and Hara, Kentaro},
  journal={Journal of Computational Physics},
  volume={508},
  pages={113030},
  year={2024},
  publisher={Elsevier}
}

@article{Boccelli2020,
	title={A 14-moment maximum-entropy description of electrons in crossed electric and magnetic fields},
	author={Boccelli, S and Giroux, F and Magin, TE and Groth, CPT and McDonald, JG},
	journal={Physics of Plasmas},
	volume={27},
	number={12},
	pages={123506},
	year={2020},
	publisher={AIP Publishing LLC}
}

@article{Mead1984,
	title={Maximum entropy in the problem of moments},
	author={Mead, Lawrence R and Papanicolaou, Nikos},
	journal={Journal of Mathematical Physics},
	volume={25},
	number={8},
	pages={2404--2417},
	year={1984},
	publisher={American Institute of Physics}
}

@article{Levermore1996,
	title={Moment closure hierarchies for kinetic theories},
	author={Levermore, C David},
	journal={Journal of Statistical Physics},
	volume={83},
	number={5},
	pages={1021--1065},
	year={1996},
	publisher={Springer}
}

@article{Bohmer2020,
	title={Entropic quadrature for moment approximations of the Boltzmann-BGK equation},
	author={B{\"o}hmer, Niclas and Torrilhon, Manuel},
	journal={Journal of Computational Physics},
	volume={401},
	pages={108992},
	year={2020},
	publisher={Elsevier}
}

@article{Sadr2021,
	title={Coupling kinetic and continuum using data-driven maximum entropy distribution},
	author={Sadr, Mohsen and Wang, Qian and Gorji, M Hossein},
	journal={Journal of Computational Physics},
	pages={110542},
	year={2021},
	publisher={Elsevier}
}

@article{McGraw1997,
	title={Description of aerosol dynamics by the quadrature method of moments},
	author={McGraw, Robert},
	journal={Aerosol Science and Technology},
	volume={27},
	number={2},
	pages={255--265},
	year={1997},
	publisher={Taylor \& Francis}
}

@article{Fox2008,
	title={A quadrature-based third-order moment method for dilute gas-particle flows},
	author={Fox, Rodney O},
	journal={Journal of Computational Physics},
	volume={227},
	number={12},
	pages={6313--6350},
	year={2008},
	publisher={Elsevier}
}

@article{Desjardins2008,
	title={A quadrature-based moment method for dilute fluid-particle flows},
	author={Desjardins, Olivier and Fox, Rodney O and Villedieu, Philippe},
	journal={Journal of Computational Physics},
	volume={227},
	number={4},
	pages={2514--2539},
	year={2008},
	publisher={Elsevier}
}

@article{Fox2018,
	title={Conditional hyperbolic quadrature method of moments for kinetic equations},
	author={Fox, Rodney O and Laurent, Fr{\'e}d{\'e}rique and Vi{\'e}, Aymeric},
	journal={Journal of Computational Physics},
	volume={365},
	pages={269--293},
	year={2018},
	publisher={Elsevier}
}

@article{vanCappellen2021,
	title={Higher order hyperbolic quadrature method of moments for solving kinetic equations},
	author={Van Cappellen, Maxim and Vetrano, Maria Rosaria and Laboureur, Delphine},
	journal={Journal of Computational Physics},
	volume={436},
	pages={110280},
	year={2021},
	publisher={Elsevier}
}

@article{Fox2021,
	title={Hyperbolic quadrature method of moments for the one-dimensional kinetic equation},
	author={Fox, Rodney and Laurent, Fr{\'e}d{\'e}rique},
    journal={SIAM Journal on Applied Mathematics},
    volume={82}, 
    number={2},
	year={2022}
}

@phdthesis{Crestetto2012,
	title = {Optimisation de m\'{e}thodes num\'{e}riques pour la physique des plasmas. Application aux faisceaux de particules charg\'{e}es.},
	author = {Crestetto, A.},
	year = {2012},
	institution = {Universit\'{e} de Strasbourg},
	note ={in French}
}

@article{Cheng2014,
	title={A class of quadrature-based moment-closure methods with application to the Vlasov--Poisson--Fokker--Planck system in the high-field limit},
	author={Cheng, Yongtao and Rossmanith, James A},
	journal={Journal of Computational and Applied Mathematics},
	volume={262},
	pages={384--398},
	year={2014},
	publisher={Elsevier}
}

@article{Berger2025,
  title={Comparison of high-order moment models for the ion dynamics in a bounded low-temperature plasma},
  author={Berger, Anatole and Lequette, N and Magin, T and Bourdon, Anne and Alvarez Laguna, A},
  journal={Physics of Plasmas},
  volume={32},
  number={10},
  year={2025},
  publisher={AIP Publishing}
}

@article{Yuan2012,
	title={An extended quadrature method of moments for population balance equations},
	author={Yuan, C and Laurent, Fr{\'e}d{\'e}rique and Fox, RO},
	journal={Journal of Aerosol Science},
	volume={51},
	pages={1--23},
	year={2012},
	publisher={Elsevier}
}

@inproceedings{Chalons2010,
	title={A multi-Gaussian quadrature method of moments for gas-particle flows in a LES framework},
	author={Chalons, C and Fox, RO and Massot, M},
	booktitle={Proceedings of the Summer Program},
	pages={347--358},
	year={2010},
	organization={Center for Turbulence Research}
}

@article{Chalons2017,
	title={Multivariate Gaussian extended quadrature method of moments for turbulent disperse multiphase flow},
	author={Chalons, Christophe and Fox, Rodney O and Laurent, Fr{\'e}d{\'e}rique and Massot, Marc and Vi{\'e}, Aymeric},
	journal={Multiscale Modeling \& Simulation},
	volume={15},
	number={4},
	pages={1553--1583},
	year={2017},
	publisher={SIAM}
}

@article{Nguyen2016,
	title={Solution of population balance equations in applications with fine particles: mathematical modeling and numerical schemes},
	author={Nguyen, Tan Trung and Laurent, Fr{\'e}d{\'e}rique and Fox, Rodney O and Massot, Marc},
	journal={Journal of Computational Physics},
	volume={325},
	pages={129--156},
	year={2016},
	publisher={Elsevier}
}

@article{Pigou2018,
	title={New developments of the extended quadrature method of moments to solve population balance equations},
	author={Pigou, Maxime and Morchain, J{\'e}r{\^o}me and Fede, Pascal and Penet, Marie-Isabelle and Laronze, Geoffrey},
	journal={Journal of Computational Physics},
	volume={365},
	pages={243--268},
	year={2018},
	publisher={Elsevier}
}

@misc{HYPRE,
	title = {{HYPRE}},
	year = {2021},
	howpublished = {\url{https://github.com/hypre-space/hypre}},
	note = {Retrieved 2021}
}

@article{Xie2023,
  title={Fusion reactivities with drift bi-Maxwellian ion velocity distributions},
  author={Xie, Huasheng and Tan, Muzhi and Luo, Di and Li, Zhi and Liu, Bing},
  journal={Plasma Physics and Controlled Fusion},
  volume={65},
  number={5},
  pages={055019},
  year={2023},
  publisher={IOP Publishing}
}

@misc{LXCAT-Phelps,
title={Phelps database},
howpublished={\url{www.lxcat.net}, retrieved on {January 6, 2026.}},
year={2026},
note={\url{http://jilawww.colorado.edu/~avp/}}
}

@misc{LXCAT-Biagi,
title={Biagi v7.1 database}, 
howpublished={\url{www.lxcat.net}, retrieved on July 24, 2025.},
year={2025},
note={Cross sections extracted from program {MAGBOLTZ}, version 7.1 {June 2004}.}
}

@misc{LXCAT-Hayashi,
title={Hayashi database}, 
howpublished={\url{www.lxcat.net}, retrieved on June 23, 2025.},
year={2025},
}

@techreport{Hayashi2003,
    author = {Hayashi, Makoto},
    title={Bibliography of electron and photon cross sections with atoms and molecules published in the 20th century. Argon},
institution = {National Institute for Fusion Science},
  year={2003},
  note={Report NIFS-DATA-72}
}

@article{Pancheshnyi2012,
  title={{The LXCat project: Electron scattering cross sections and swarm parameters for low temperature plasma modeling}},
  author={Pancheshnyi, Sergey and Biagi, S and Bordage, Marie-Claude and Hagelaar, GJM and Morgan, WL and Phelps, AV and Pitchford, Leanne C},
  journal={Chemical Physics},
  volume={398},
  pages={148--153},
  year={2012},
  publisher={Elsevier}
}

@article{Pitchford2017,
  title={{LXcat: An open-access, web-based platform for data needed for modeling low temperature plasmas}},
  author={Pitchford, Leanne C and Alves, Luis L and Bartschat, Klaus and Biagi, Stephen F and Bordage, Marie-Claude and Bray, Igor and Brion, Chris E and Brunger, Michael J and Campbell, Laurence and Chachereau, Alise and others},
  journal={Plasma Processes and Polymers},
  volume={14},
  number={1-2},
  pages={1600098},
  year={2017},
  publisher={Wiley Online Library}
}

@article{Powis2024,
  title={Accuracy of the explicit energy-conserving particle-in-cell method for under-resolved simulations of capacitively coupled plasma discharges},
  author={Powis, Andrew T and Kaganovich, Igor D},
  journal={Physics of Plasmas},
  volume={31},
  number={2},
  year={2024},
  publisher={AIP Publishing}
}

@book{Szego,
title={Orthogonal polynomials, 4th edition},
author={Szeg\"{o}, Gabor},
year={1975},
publisher={American Mathematical Society},
note={{Theorem 3.2.1, p.42}}
}

\end{document}